\documentclass{article}

\pdfoutput=1 
\usepackage{jheppub}
\usepackage{graphicx,verbatim}
\usepackage{psfrag, times}
\usepackage[T1]{fontenc}
\usepackage{epsf}
\usepackage{mathtools}

\usepackage{amsmath}
\usepackage{amsfonts}
\usepackage{amssymb}
\usepackage{mathrsfs}
\usepackage{slashed}
\usepackage{xcolor}

\def\be{\begin{equation}}
\def\ee{\end{equation}}
\def\bc{\begin{center}}
\def\ec{\end{center}}

\def\k{\kappa}
\def\bR{{\mathbb{R}}}

\def\bbC{{\mathbb{C}}}

\def\bbX{{\mathbb{X}}}

\def\bbR{{\mathbb{R}}}
\def\bA{\bar{A}}

\def\bx{{\mathbf{x}}}
\def\by{{\mathbf{y}}}

\def\cN{{\mathcal{N}}}

\def\cT{{\mathcal{T}}}

\def\cU{{\mathcal{U}}}

\def\cW{{\mathcal{W}}}

\def\cZ{{\mathcal{Z}}}

\def\r2{{\sqrt{2}}}

\def\bea{\begin{eqnarray}}
\def\eea{\end{eqnarray}}

\def\cT{{\mathcal{T}}}

\def\bK{{\bf{K}}}

\def\bc{{\bf{c}}}

\def\bbZ{{\mathbb{Z}}}
\def\bN{{\bf{N}}}

\def\rx{{\rm{x}}}
\def\ry{{\rm{y}}}

\def\fJ{{\mathfrak{J}}}
\def\bi{{\mathbf{i}}}
\def\bj{{\mathbf{j}}}
\def\bl{{\mathbf{l}}}
\def\bt{{\mathbf{t}}}
\def\eps{\epsilon}

\usepackage{url}
\usepackage{hyperref}
\usepackage[mathscr]{euscript}
\usepackage{CJKutf8}

\title{Quantum Elliptic Calogero-Moser Systems from Gauge Origami}
\author[]{Heng-Yu Chen${}^{1}$, Taro Kimura${}^{2}$ and Norton Lee${}^{3}$}
\affiliation{$^1$\rm Department of Physics, National Taiwan University, Taipei 10617, Taiwan}
\affiliation{$^2$\rm Department of Physics, Keio University, Kanagawa 223-8521, Japan\\
$~~$ Institut de Math\'ematiques de Bourgogne, Universit\'e Bourgogne Franche-Comt\'e, 21078 Dijon, France}
\affiliation{$^3$\rm C. N. Yang Institute for Theoretical Physics, Stony Brook University, Stony Brook, NY 11794, USA}
\emailAdd{ heng.yu.chen@phys.ntu.edu.tw}
\emailAdd{taro.kimura@u-bourgogne.fr}
\emailAdd{norton.lee@stonybrook.edu}

\vspace{2cm}
\abstract{We systematically study the interesting relations between the quantum elliptic Calogero-Moser system (eCM) and its generalization, and their corresponding supersymmetric gauge theories. In particular, we construct the suitable characteristic polynomial for the eCM system by considering certain orbifolded instanton partition function of the corresponding gauge theory. This is equivalent to the introduction of certain co-dimension two defects. We next generalize our construction to the folded instanton partition function obtained through the so-called ``gauge origami'' construction and precisely obtain the corresponding characteristic polynomial for the doubled version, named the elliptic double Calogero-Moser (edCM) system.}

\begin{document}

\maketitle

\section{Introduction and Summary}
\paragraph{}
Among the plethora of integrable systems, the elliptic Calogero-Moser types (eCM) have continuously fascinated mathematicians and physicists (see \cite{Ruijsenaars:1999PF} for a good introduction). Of our particular interests, it is well-known that the classical spectral curve of the eCM integrable system associated with Lie algebra ${\rm Lie}(G)$ can be directly identified with the Seiberg-Witten curve of four dimensional $\cN=2^*$ theory with gauge group $G$ \cite{Gorsky:1994dj, Nekrasov:1995nq, Martinec:1995by}, see also \cite{DW1, DPLecture}.
Indeed, this correspondence serves as one of the earliest examples demonstrating the close connections between certain integrable systems and the gauge theories with eight supersymmetries. Moreover with the tremendous advances in the localization computations for supersymmetric partition functions and other protected observables (see \cite{Pestun:2016zxk} for extensive reviews), we can extend the correspondence to the quantum level. In general, there can be multiple deformation parameters $\{\epsilon_a\}$ in these localization computations depending on the dimensionality of the  supersymmetric gauge theories considered. It was proposed that $\{\epsilon_a\}$ can be identified in general with the Planck constants when quantizing the original classical integrable systems, such that turning off one or more of $\{\eps_a\}$ can be interpreted as recovering certain semi-classical limit \cite{Nikita-Shatashvili}. 
In such a limit, the Bethe Ansatz equation (BAE) of the quantum integrable system can be recovered from the saddle point equation of the corresponding gauge theory partition function computed by localization techniques. In the full quantum case with all $\{\epsilon_a\}$ kept finite, the instanton partition function can be identified as the eigenfunction of quantum integrable system's Hamiltonian. Such a quantity is computed by using the so called qq-characters~\cite{Nikita:I,Nikita:V}. 
\paragraph{}
One of the hallmarks of the integrability in a dynamical system is the existence of the commuting Hamiltonians, the generating function of them is a finite degree polynomial in the appropriate spectral parameter, known as the ``characteristic polynomial'' or ``transfer matrix''. The gauge theoretic counterpart of the characteristic polynomial has been shown in certain cases to be the generating function of chiral rings, such as $\cN=2$ SQCD/XXX-spin chain and its linear quiver generalization~\cite{Gorsky:1995zq,Martinec:1995by,Seiberg:1996nz,Gorsky:1996hs}. It is also very natural to consider the quantum version of this story in the same vein as discussed in the previous paragraph, and identify the commuting quantum Hamiltonians with the chiral ring operators~\cite{Nikita-Shatashvili,Dorey:2011pa,HYC:2011}.  
However it is unsatisfactory that eCM systems and their corresponding gauge theories have somehow evaded this general line of developments. As we will review later in Section \ref{eCM-N=2*}, even though we can readily recover the BAE for the eCM system through the saddle point analysis of partition function~\cite{Nikita-Shatashvili}, a naive gauge theoretic construction of characteristic polynomial however failed to yield the correct commuting quantum Hamiltonians. The situation can be rectified by constructing a certain regular function of spectral parameter with the appropriate degree, which will be named $\bbX$-function.%
\footnote{%
This $\bbX$-function itself is also known as the fundamental $q$-character of $\widehat{A}_0$ quiver constructed in~\cite{Nikita-Pestun-Shatashvili}.
See also~\cite{FJMM} for another construction through the quantum toroidal algebra of $\mathfrak{gl}_1$.
As mentioned in this paper, we need to consider the orbifolded version of the $\bbX$-function in order to extract the commuting Hamiltonians of the eCM system.
}
More precisely this construction is a two step process as we will discuss in Section \ref{X-function}. First, we will introduce the co-dimension two surface defects into the gauge theory through the orbifolding~\cite{Feigin:2011SM,Finkelberg:2010JEMS,K-T,Nikita:IV}, this also has the effect of splitting the original gauge theory into multiple orbifolded copies.
The $\bbX$-function then arises from summing over a suitable instanton partition function for each orbifolded copy. We will demonstrate that the commuting Hamiltonians of the eCM system can indeed be extracted from the resultant $\bbX$-function. 
\paragraph{}
We next apply our story to an inherently quantum generalization of the eCM system, known as the elliptic double Calogero-Moser system (edCM)~\cite{Nikita:V}.%
\footnote{The trigonometric version is studied, e.g., in~\cite{SV,HiJack}.}
 This implies its corresponding gauge theory is necessarily well-defined only when at least one of  $\{\eps_i\}$ is  turned on. Indeed, the consistent gauge theoretic construction related to the edCM system is known as ``Gauge Origami''~\cite{Nikita:III}.%
 \footnote{\begin{CJK}{UTF8}{ipxm}ゲージ折紙~(日本語)\end{CJK}; \begin{CJK}{UTF8}{bsmi}規範摺紙~(中文 繁體字)\end{CJK}; \begin{CJK}{UTF8}{gbsn}规范折纸~(中文 简体字)\end{CJK}.}
This arises from the intersecting D-brane configuration in the presence of background fluxes, which corresponds to turning on $\{\epsilon_i\}$ \cite{NaveenNikita}, as we will review this in Section~\ref{GaugeOrigami}.  In Section~\ref{eDCM-Origami}, we will explicitly construct the resultant instanton partition functions, derive the possible BAE from its saddle point equation, and follow our earlier procedures for the eCM system to construct the $\bbX$-function in this case. Finally we demonstrate the validity of $\bbX$-function by recovering the correct commuting Hamiltonians which are expressed in terms of the Dunkl operators generalized to the edCM system.
We should comment here that the connection between edCM systems and the so-called ``folded instanton'' configuration derived from gauge origami construction was noticed in \cite{Nikita:V}, in this work we firmly established this connection by working out the relevant details in steps.
\paragraph{}
We discuss various future directions in Section~\ref{sec:discussion}.
We relegate our various definitions of functions and some of the computational details in a series of Appendices.

\section{Elliptic Calogero-Moser Model and $\mathcal{N}=2^*$ Theory}\label{eCM-N=2*}
\paragraph{}
It is well known that the elliptic Calogero-Moser (eCM) model (see \cite{Ruijsenaars:1999PF} for an excellent review), which is an one-dimensional quantum mechanical system of $N$ particles with Hamiltonian of the form:
\begin{equation}\label{H-CS}
	\hat{H}_{\rm eCM}=-\frac{\hbar^2}{2}\sum_{\alpha=1}^N\frac{\partial^2}{\partial \rx_\alpha^2}+m(m+\hbar)\sum_{1\leq \alpha<\beta  \leq N}\wp(\rx_\alpha-\rx_\beta),
\end{equation}
is closely related to four dimensional $\mathcal{N}=2^*$ $SU(N)$ gauge theory.%
\footnote{We choose this notation intentionally. We will identify adjoint mass $m$ of $\mathcal{N}=2^*$ with potential coupling of Calogero-Moser system in the end of Section~\ref{X-function}.}
Here the interacting potential is given in terms of Weierstrass $\wp(u)$-function  defined in eq.~\eqref{Def:p-function}.
\paragraph{}
When $\frac{\hbar}{m} \to 0$, \eqref{H-CS} approaches its classical limit, 
\be
H_{\rm eCM}=\frac{1}{2}\sum_{\alpha=1}^N p_\alpha^2+m^2\sum_{1\leq \alpha<\beta\leq N}\wp(\rx_\alpha-\rx_\beta).
\ee
It has been proven that this system encodes the underlying classical integrable structure of $\mathcal{N}=2^*$ super Yang-Mills theory by identifying its spectral curve with the gauge theory Seiberg-Witten curve in many early literature such as \cite{DP1, DW1} and see \cite{DPLecture} for a more complete list of references.
In this note, we aim to extend in several directions the quantum version of such a correspondence from various new results in gauge theories.

\subsection{Bethe Ansatz Equation from Instanton Partition Function}
\label{sec:BAEfromInstanton}
\paragraph{}
As a warm up example setting up our subsequent notations and terminologies, as well as illustrating the problem, 
we first recall how the BAE of the eCM model can arise from the instanton partition function of $\mathcal{N}=2^*$ $SU(N)$ gauge theory. The instanton partition function can be obtained from the localization computation in $\Omega$-background and is expressed in terms of a summation over all allowed instanton configurations \cite{Nekrasov:2002qd,NO1}, each of them is labeled by a set of $N$ Young diagrams $\vec{\lambda}=(\lambda^{(1)},\dots,\lambda^{(N)})$. Each Young diagram $\lambda^{(\alpha)}$ for $\alpha = 1, \ldots, N$ is labeled by row vectors: $\lambda^{(\alpha)}=(\lambda^{(\alpha)}_1,\lambda^{(\alpha)}_2,\dots)$ with non-negative entries such that:
\begin{equation}
	\lambda_i^{(\alpha)}\geq\lambda^{(\alpha)}_{i+1}, \quad i = 1, 2, \dots,
\end{equation}
which denote the number of box of each row. 
Let us define the following parameters:
\begin{equation}\label{Def:xx0}
	x_{\alpha i}=a_\alpha+(i-1)\epsilon_1+\lambda_i^{(\alpha)}\epsilon_2,\qquad
	x_{\alpha i}^{(0)}=a_\alpha+(i-1)\epsilon_1,
\end{equation}
where $(\eps_1, \eps_2)$ are the $\Omega$ background deformation parameters. 
The instanton partition function is now written as the summation:
\begin{subequations}\label{N=2* IP}
\begin{align}
\mathcal{Z}_{\text{inst}}&=\sum_{\{\vec{\lambda}\}}\mathfrak{q}^{|\vec{\lambda}|} \mathcal{Z}_{\rm inst}[\vec{\lambda}],\\
	\mathcal{Z}_\text{inst}[\vec{\lambda}]&=\prod_{(\alpha i)\neq(\beta j)} \frac{\Gamma(\epsilon_2^{-1}(x_{\alpha i}-x_{\beta j}-\epsilon_1))}{\Gamma(\epsilon_2^{-1}(x_{\alpha i}-x_{\beta j}))}\cdot\frac{\Gamma(\epsilon_2^{-1}(x_{\alpha i}^{(0)}-x_{\beta j}^{(0)}))}{\Gamma(\epsilon_2^{-1}(x_{\alpha i}^{(0)}-x_{\beta j}^{(0)}-\epsilon_1))} \nonumber\\
	&  \qquad\times\frac{\Gamma(\epsilon_2^{-1}(x_{\alpha i}-x_{\beta j}-m))}{\Gamma(\epsilon_2^{-1}(x_{\alpha i}-x_{\beta j}-m-\epsilon_1))}\cdot\frac{\Gamma(\epsilon_2^{-1}(x_{\alpha i}^{(0)}-x_{\beta j}^{(0)}-m-\epsilon_1))}{\Gamma(\epsilon_2^{-1}(x_{\alpha i}^{(0)}-x_{\beta j}^{(0)}-m))},
\end{align}
\end{subequations}
with 
\begin{align}
\mathfrak{q}=e^{2\pi i\tau}
\label{eq:coupling_const}
\end{align}
where $\tau$ is the complexified gauge coupling, and $m$ is the complex adjoint mass. 
\paragraph{}
Let us consider the so-called Nekrasov-Shatashvili limit (or NS limit for short) \cite{Nikita-Shatashvili}, such that $\epsilon_2\to0$ with $\epsilon_1=:\epsilon$ fixed, 
and take the Stirling approximation on $\Gamma$-function, we obtain:
\begin{align}
\lim_{\eps_2 \to 0}\mathcal{Z}_{\rm inst}[\vec{\lambda}]=\exp\left[\frac{1}{2\epsilon_2}\right.\sum_{(\alpha i)\neq(\beta j)}
&f((x_{\alpha i}-x_{\beta j}-\epsilon)-f((x_{\alpha i}-x_{\beta j}+\epsilon) \nonumber\\
& -f(x_{\alpha i}^{(0)}-x_{\beta j}^{(0)}-\epsilon)+f(x_{\alpha i}^{(0)}-x_{\beta j}^{(0)}+\epsilon) \nonumber\\
&+f(x_{\alpha i}-x_{\beta j}-m)-f(x_{\alpha i}-x_{\beta j}+m) \nonumber\\
& -f(x_{\alpha i}^{(0)}-x_{\beta j}^{(0)}-m)+f(x_{\alpha i}^{(0)}-x_{\beta j}^{(0)}-m)\nonumber \\
&-f(x_{\alpha i}-x_{\beta j}-m-\epsilon)+f(x_{\alpha i}-x_{\beta j}+m+\epsilon) \nonumber\\
& +\left.f(x_{\alpha i}^{(0)}-x_{\beta j}^{(0)}-m-\epsilon)-f(x_{\alpha i}^{(0)}-x_{\beta j}^{(0)}-m+\epsilon)\right] ,
\end{align}
where $f(x) = x(\log x-1)$.
In this limit, the combination $\epsilon_2 \lambda^{(\alpha)}_i$ becomes continuous, such that the sum over the discrete Young diagrams can be approximated by a continuous integral over a set of infinite integration variables $\{x_{\alpha i}\}$,
\begin{equation}
\lim_{\epsilon_2 \to 0} \mathcal{Z}_\text{inst}=\int\prod_{(\alpha i)}dx_{\alpha i}\exp \left[\frac{1}{\epsilon_2}\mathcal{H}_{\text{inst}}(x_{\alpha i})\right].
\end{equation}
The instanton functional $\mathcal{H}_{\text{inst}}(x_{\alpha i})$ takes the form of:
\begin{equation}
\mathcal{H}_{\text{inst}}(x_{\alpha i})=\mathcal{U}(x_{\alpha i})-\mathcal{U}(x^{(0)}_{\alpha i}),
\end{equation}
where we have also defined:
\begin{equation}\label{A1U}
\begin{aligned}
\mathcal{U}(x_{\alpha i})=\log\mathfrak{q}\sum_{(\alpha i)}x_{\alpha i}+\frac{1}{2}\sum_{(\alpha i)\neq(\beta j)}
&\{ f(x_{\alpha i}-x_{\beta j}-\epsilon)-f(x_{\alpha i}-x_{\beta j}+\epsilon) \\
& +f(x_{\alpha i}-x_{\beta j}-m)-f(x_{\alpha i}-x_{\beta j}+m) \\
&-f(x_{\alpha i}-x_{\beta j}-m-\epsilon)+f(x_{\alpha i}-x_{\beta j}+m+\epsilon)\}.
\end{aligned}
\end{equation}
Here we have introduced the instanton density $\rho(x)$ which is a non-vanishing constant along $\mathfrak{J}=\bigcup_{\alpha i}[x_{\alpha i}^{(0)},x_{\alpha i}]$ and zero everywhere else to rewrite $\mathcal{H}_{\text{inst}}$. Furthermore we can define the combinations:
\begin{equation}
{R}(x)=\frac{P(x-m)P(x+m+\epsilon)}{P(x)P(x+\epsilon)};\quad{G}(x)=\frac{d}{dx}\log\frac{(x+m+\epsilon)(x-m)(x-\epsilon)}{(x-m-\epsilon)(x+m)(x+\epsilon)},
\end{equation}
where $P(x)=\prod_{\alpha=1}^N(x-a_\alpha)$. Together, the instanton partition functional $\mathcal{H}_{\text{inst}}$ can be written as:
\begin{equation}
\mathcal{H}_{\text{inst}}(x_{\alpha i})=-\frac{1}{2}\operatorname{PV}\int_{\fJ\times \fJ} dxdy\rho(x){G}(x-y)\rho(y)+\int_\fJ dx\rho(x)\log\mathfrak{q}{R}(x),
\end{equation}
where the symbol PV means the principal value integral.
In $\epsilon_2\to0$ limit, the integration should be dominated by saddle point configurations, which yield:
\begin{equation}\label{varHA0}
\frac{\delta\mathcal{H}_{\text{inst}}[\rho]}{\delta x_{\alpha i}}=-\int_{\mathfrak{J}}dy{G}(x_{\alpha i}-y)\rho(y)+\log(\mathfrak{q}{R}(x_{\alpha i}))=0.
\end{equation}
As ${G}(x)$ is a total derivative, one obtains
\begin{equation}\label{A0T}
\begin{aligned}
1
=-\mathfrak{q}\frac{{Q}(x_{\alpha i}+m+\epsilon){Q}(x_{\alpha i}-m){Q}(x_{\alpha i}-\epsilon)}{{Q}(x_{\alpha i}-m-\epsilon){Q}(x_{\alpha i}+m){Q}(x_{\alpha i}+\epsilon)},
\end{aligned}
\end{equation}
where 
\begin{equation}
Q(x)=\prod_{\alpha=1}^N\prod_{j=1}^\infty(x-x_{\alpha j}).
\end{equation}
We often call the Young diagram $\vec{\lambda}_*$ satisfying eq.~\eqref{A0T} the "Limit shape configuration," which dominates the summation in eq.~\eqref{N=2* IP} under NS-limit:
\begin{align}\label{limit shape IP}
    \mathcal{Z}_\text{inst}\approx \mathcal{Z}_\text{inst}[\vec{\lambda}_*].
\end{align}

\paragraph{}
To see how BAE of the quantum eCM model emerges, we consider the twisted superpotential arising from the full partition function $\mathcal{Z}_{\widehat{A}_0}$: $\mathcal{W}_{\widehat{A}_0}=\lim_{\epsilon_2\to0}[\epsilon_2\log\mathcal{Z}_{\widehat{A}_0}]=\mathcal{W}_\text{classical}+\mathcal{W}_{\text{1-loop}}+\mathcal{W}_{\text{inst}}$. 
For the non-perturbative part we have:
\begin{equation}
\mathcal{W}_{\text{inst}}(a_\alpha)=\mathcal{H}_{\text{inst}}(x_{\alpha i})=\mathcal{U}(x_{\alpha i})-\mathcal{U}(x_{\alpha i}^{(0)}),
\end{equation}
where $\cU(x)$ is defined in eq.~\eqref{A1U}. 
While the remaining classical twisted superpotential is
\begin{equation}
	\cW_\text{classical}(a_\alpha)=-\log\mathfrak{q}\sum_{\alpha=1}^N\frac{a_\alpha^2}{2\epsilon},
\end{equation}
and the perturbative one-loop twisted superpotential is
\begin{align}
 \cW_\text{1-loop}(a_\alpha) = \mathcal{U}(x_{\alpha i}^{(0)}) - \log \mathfrak{q} \sum_{(\alpha i)} x_{\alpha i}^{(0)}.
\end{align}
Unlike the gauge theories with massive fundamental hypermultiplets \cite{HYC:2011}, there is no natural truncation condition  on the Young diagrams labeling instanton partition function.
Instead, the equation of motion for the functional $\cW_{\widehat{A}_0}$ is given by:
\begin{equation}
\frac{1}{2\pi i}\frac{\partial {\cW_{\widehat{A}_0}}(a_{\alpha})}{\partial a_{\alpha}}=n_{\alpha};\qquad n_\alpha\in\mathbb{Z},
\end{equation}
explicitly one obtains:
\begin{equation}\label{BAE-eCM}
	-\frac{a_\alpha}{\epsilon}\log\mathfrak{q}+\sum_{\beta\neq\alpha}\log\frac{\Gamma\left(\frac{a_\alpha-a_\beta}{\epsilon}\right)}{\Gamma\left(-\frac{a_\alpha-a_\beta}{\epsilon}\right)}\frac{\Gamma\left(\frac{-m-(a_\alpha-a_\beta)}{\epsilon}\right)}{\Gamma\left(\frac{-m+a_\alpha-a_\beta}{\epsilon}\right)}=2\pi in_\alpha,
\end{equation}
using the following identity:
\begin{align}\label{How to Gamma}	
\frac{\partial}{\partial a_\alpha}\frac{1}{2}\sum_{(\alpha i)\neq(\beta j)} \left(f(x_{\alpha i}-x_{ \beta j}-\epsilon)-f(x_{\alpha i}-x_{\beta j}+\epsilon)\right) =\sum_{\beta\neq\alpha}\log\frac{\Gamma\left(\frac{a_\alpha-a_\beta}{\epsilon}\right)}{\Gamma\left(-\frac{a_\alpha-a_\beta}{\epsilon}\right)}.
\end{align}
Exponentiating both sides of equation \eqref{BAE-eCM} gives
\begin{equation}\label{BAEA0}
1=\mathfrak{q}^{-\frac{a_\alpha}{\epsilon}}\prod_{\beta\neq\alpha}\frac{\Gamma\left(\frac{a_\alpha-a_\beta}{\epsilon}\right)}{\Gamma\left(-\frac{a_\alpha-a_\beta}{\epsilon}\right)}\frac{\Gamma\left(\frac{-m-(a_\alpha-a_\beta)}{\epsilon}\right)}{\Gamma\left(\frac{-m+a_\alpha-a_\beta}{\epsilon}\right)},
\end{equation}
this is the BAE of the eCM system~\cite{Nikita-Shatashvili}.
\paragraph{}
Here we would like to introduce the following $T(x)$-function:
\begin{equation}\label{TA0}
T(x)=\frac{{Q}(x+\epsilon)}{{Q}(x)}\left[1+\mathfrak{q}\frac{{Q}(x+m+\epsilon){Q}(x-m){Q}(x-\epsilon)}{{Q}(x+m){Q}(x-m-\epsilon){Q}(x+\epsilon)}\right],
\end{equation} 
which will be proposed as a tentative characteristic polynomial for generating the commuting Hamiltonians of the eCM system later.
We can also recast $T(x)$ in a more illuminating form by defining:
\begin{equation}\label{Y-func}
Y(x)=\frac{Q(x)}{Q(x-\epsilon)},
\end{equation}
and rewrite $T(x)$ as
\begin{equation}\label{firstT}
T(x)={Y(x+\epsilon)}\left[1+\mathfrak{q}\frac{Y(x-m)Y(x+m+\epsilon)}{Y(x)Y(x+\epsilon)}\right].
\end{equation}
While similar $T(x)$ works as the characteristic polynomial for XXX spin chain arising from superconformal QCD, see e.g.~\cite{HYC:2011},
we will see momentarily that in order to correctly reproduce the commuting Hamiltonians for the eCM system, we need to further enhance $T(x)$ as defined in \eqref{firstT}.

\subsection{Finding the commuting Hamiltonians: Warm Up}
\paragraph{}
After demonstrating how the BAE can arise from four dimensional $\cN=2^*$ $SU(N)$ gauge theory, it is natural to consider if it is possible to similarly obtain the commuting Hamiltonians.
To this end, we first consider the
$T(x)$ function defined in \eqref{A0T} and \eqref{firstT} which is a natural candidate for generating all commuting Hamiltonians of the eCM system by identifying its expansion coefficients in appropriate asymptotic regime. 
\subsubsection{T-function from gauge theory}
As a simplification to illustrate the procedures, let us consider pure $\mathcal{N}=2$ $SU(N)$ gauge theory by setting $m \to \infty$ to integrate out the adjoint hypermultiplet,
it is known that the associated integrable system is $\widehat{A}_{N-1}$ Toda lattice (\begin{CJK}{UTF8}{ipxm}戸田格子\end{CJK}) system:
\begin{equation}\label{H-Toda}
 \hat{H}_{\rm Toda}= - \frac{\hbar^2}{2} \sum_{\alpha=1}^N \frac{\partial^2}{\partial \rx_\alpha^2} 
  +\Lambda^2\sum_{\alpha=1}^{N}e^{\rx_\alpha-\rx_{\alpha-1}};\quad \rx_{\alpha}\sim \rx_{\alpha+N}, \quad \mathfrak{q}=\Lambda^{2N}.
\end{equation}
Now in the $m\to\infty$ limit, the saddle point equation \eqref{A0T} now becomes {BAE for $\widehat{A}_{N-1}$-Toda system:}
\begin{equation}\label{A1}
1=-\mathfrak{q}\frac{Q(x_{\alpha i}-\epsilon)}{Q(x_{\alpha i}+\epsilon)},
\end{equation}
such that $T(x)$ defined in \eqref{firstT} reduces to:
\begin{equation}\label{Tfunc}
T(x)=\frac{Q(x+\epsilon)}{Q(x)}\left[1+\mathfrak{q}\frac{Q(x)}{Q(x-\epsilon)}\right]=Y(x+\epsilon)+\mathfrak{q}\frac{1}{Y(x)},
\end{equation}
as can also be deduced from $\lim_{m \to \infty}Y(x\pm m) \to 1$.
\paragraph{}
We will now show that $T(x)$ is a degree $N$ polynomial in spectral parameter $x$. Using \eqref{A1}, we can see that the apparent poles of $T(x)$ coming from poles of $Y(x+\epsilon)$ are canceled by the corresponding zeros in the bracket. This proves that $T(x)$ is analytic in the complex $x$-plane (excluding $x=\infty$). 
To prove $T(x)$ has the correct degree, we first consider large $x$ behavior of $Y(x)$. When $x$ is large, we may approximate $x_{\alpha i}\approx x_{\alpha i}^{(0)}$. Thus the asymptotic behavior of $Y(x)$ behaves as:
\begin{equation}
 Y(x)\sim\prod_{\alpha=1}^N\prod_{i=1}^\infty\frac{x-a_\alpha-(i-1)\epsilon}{x-a_{\alpha}-(i-1)\epsilon - \epsilon}=\prod_{\alpha=1}^N(x-a_\alpha)\sim x^N
  \quad \text{at} \quad x \to \infty.
\end{equation}
We conclude that \eqref{Tfunc} constructed based on saddle point equation is a degree $N$ polynomial. 
Next we would like to see if it can be directly related to the characteristic polynomial of Toda lattice by checking if we can recover Hamiltonian given in \eqref{H-Toda}.
\subsubsection{Surface defect via orbifolding}
Here we would like to introduce a co-dimension two surface defect on $\mathbb{C}_{1}\subset\mathbb{R}^4=\mathbb{C}_1\times\mathbb{C}_2$, with $\mathbb{Z}_N$ orbifolding on the coordinates of $\mathbb{C}^2$ by $(\mathbf{z}_1,\mathbf{z}_2)\to(\mathbf{z}_1,\zeta \mathbf{z}_2)$ where $\zeta^N=1$. This orbifolding procedure commutes with NS limit\footnote{The introduction of full surface defects into pure $\cN=2$ SYM and $\cN=2^*$ theories, and their connections with quantum integrable systems was also considered earlier in \cite{Nawata:2014nca}.}. We will restore $\epsilon_2$ dependence for a moment, and take NS limit after orbifolding. {Such orbifolding maps the original gauge theory (which can be considered as $A_1$ quiver) into so called handsaw quiver structure \cite{Nakajima:2011yq}.}
Shown in \cite{Nikita:V}, the instanton partition function with surface defect inserted is an eigenfunction of Hamiltonian for both Calogero-Moser and Toda cases (eq.~\eqref{H-CS} for Calogero-Moser and eq.~\eqref{H-Toda} for Toda). On the other hand, $\cN=2$ with fundamental hypermultiplet does not require such a orbifolding procedure, as discussed in \cite{HYC:2011}. 
It should be noted however that it is possible to relate these two different types of surface defects via the brane creation process similar to Hanany-Witten transition in M-theory, as discussed in \cite{Frenkel:2015rda,Jeong:2018qpc}.
\paragraph{}
A $\mathbb{Z}_N$ type surface defect in a $U(N)$ gauge theory \cite{Nikita:IV} can be characterized by a coloring function: $c:\{\alpha=1,2,\dots,N\}\to\mathbb{Z}_N$, which assigns a color $\alpha$ labeling Coulomb parameter $a_\alpha$ to an irreducible representation $R_\omega$ of $\mathbb{Z}_N$, $\omega=0,1,\dots,N-1$.
Here we choose the simplest form of $c$:
\begin{equation}
c(\alpha)=\alpha-1,
\end{equation}
this implies we can assign the Coulomb parameter $a_\alpha$ to the representation $R_{\alpha-1}$ of $\mathbb{Z}_N$.  
One may take other form of coloring function $c$ if needed. 
In principle, one can consider the lower degree orbifolding as the quotient by $\mathbb{Z}_{n < N}$. The defect corresponding to $\mathbb{Z}_N$ is called the full-type surface defect, which is relevant to our purpose. More detailed discussions can be found in~\cite{Feigin:2011SM,Finkelberg:2010JEMS,K-T,Nikita:IV}. Under $\bbZ_N$ orbifolding, the complex coupling $\mathfrak{q}$ splits into $N$ copies:
\be
\mathfrak{q}=\mathfrak{q}_0\mathfrak{q}_1\cdots\mathfrak{q}_{N-1};\quad \mathfrak{q}_{\omega+N}=\mathfrak{q}_\omega,
\ee
each $\mathfrak{q}_\omega$ is assigned to the representation $R_\omega$ of $\mathbb{Z}_N$ for complex gauge coupling. 
Under Orbifolding, counting in the instanton partition function becomes
\begin{align}\label{split q}
    \mathfrak{q}^{|\vec{\lambda}|}=\prod_{\omega}\mathfrak{q}_\omega^{k_\omega}
\end{align}
with the definition
\begin{equation}\label{K}
K_\omega:=\{(\alpha,(i,j)) \mid \alpha=1,\dots,N;\quad(i,j)\in\lambda^{(\alpha)};\quad\alpha+j-1\equiv\omega \ \text{mod} \ N \}
\end{equation}
and the following definition of notations: 
\begin{equation}\label{kv-data}
\begin{aligned}
& k_\omega=|K_\omega|,\qquad\nu_\omega=k_\omega-k_{\omega+1}.
\end{aligned}
\end{equation}
\paragraph{}
We will show how the introduction of full surface defect affects $T(x)$. Starting from its building block $Y(x)$, under orbifolding, $Y(x)$ becomes $Y(x)=\prod_{\omega}Y_\omega(x)$ where each orbifold copy is:
\begin{align}\label{Y_w}
Y_\omega(x)
=(x-a_{\omega})&\prod_{(\alpha,(i,j))\in K_{\omega}}
\left[\frac{(x-a_\alpha-(i-1)\epsilon_1-(j-1)\epsilon_2-\epsilon_1)}{(x-a_\alpha-(i-1)\epsilon_1-(j-1)\epsilon_2)}\right] \nonumber\\
&\times\prod_{(\alpha,(i,j))\in K_{\omega+1}} 
\left[\frac{(x-a_\alpha-(i-1)\epsilon_1-(j-1)\epsilon_2-\epsilon_2)}{(x-a_\alpha-(i-1)\epsilon_1-(j-1)\epsilon_2-\epsilon_2-\epsilon_1)}\right],
\end{align}
For more general coloring function, the condition in \eqref{K} should be $c(\alpha) + j \equiv \omega$. 
$K_\omega$ is a collection of Young diagram boxes which are assigned to the representation $R_\omega$ under orbifolding. Remember that orbifolding is imposed in $\mathbb{C}_2$, with $\epsilon_2$ charged. Adding or subtracting an $\epsilon_2$ moves the representation by $\pm1$. For instance if  $a_\alpha$ is assigned to representation $R_{\alpha-1}$, then $a_\alpha+\epsilon_2$ is of representation $R_{\alpha}$. This essentially splits boxes in Young diagram into different collections of $K_\omega$ based on its location as in \eqref{K}. 
Each Young diagram box in $K_\omega$ is labeled by orbifolded instanton counting parameter $\mathfrak{q}_\omega$. For each $R_\omega$, we now have orbifolded $T$-functions as
\begin{equation}
\begin{aligned}
T_\omega(x)
& =Y_{\omega+1}(x+\epsilon_+)+\frac{\mathfrak{q}_\omega }{Y_\omega(x)} ;\qquad T(x)=\prod_{\omega=0}^{N-1} T_\omega(x).\\
\end{aligned}
\end{equation}
The presence of such a surface defect is necessary for both Calogero-Moser and Toda cases for their instanton partition function to become eigenfunctions, detailed discussions can be found in \cite{Nikita:V}. 
\subsubsection{NS limit under orbifolding}
Now we resume to take the NS limit and further consider the large $x$ asymptotic of $T_{\omega}(x)$, each individual copy $Y_{\omega}(x)$ becomes:
\begin{equation}
Y_\omega(x)={(x-a_{\omega})}\exp\left[\frac{\epsilon}{x}\nu_{\omega-1}+\frac{\epsilon}{x^2}D_{\omega-1}+\cdots\right],
\end{equation}
with the definition in \eqref{K}, \eqref{kv-data}, and  
\begin{align}
    \sigma_\omega=\frac{\epsilon}{2}k_\omega+\sum_{(\alpha,(i,j))\in K_\omega}(a_\alpha+(i-1)\epsilon); \quad D_\omega=\sigma_\omega-\sigma_{\omega+1}.
\end{align}
Together with its inverse, we have:
\begin{equation}\label{T-x-large}
T_\omega(x)=x+\epsilon-a_{\omega+1}+\epsilon\nu_{\omega}+\frac{1}{x}\left[\frac{1}{2}\epsilon^2\nu_{\omega}^2-\epsilon\nu_{\omega}a_{\omega+1}+\epsilon D_\omega-\epsilon^2\nu_\omega+\mathfrak{q}_\omega \right]+\cdots,
\end{equation}
Multiplying together all the orbifold copies, we obtain that:
\begin{equation}
\begin{aligned}
T(x)
& = \prod_{\omega=0}^{N-1} T_\omega(x) \\
& = x^N+\left(\sum_\omega \epsilon-a_{\omega+1}+\epsilon\nu_{\omega}\right)x^{N-1} \\
& \quad +\left[\sum_{\omega<\omega'} (\epsilon-a_{\omega+1}+\epsilon\nu_{\omega}) (\epsilon-a_{\omega'+1}+\epsilon\nu_{\omega'})+\sum_{\omega}\frac{1}{2}\epsilon^2\nu_{\omega}^2-\epsilon\nu_{\omega}a_{\omega+1}+\epsilon_1D_\omega-\epsilon^2\nu_\omega+\mathfrak{q}_\omega\right]+\dots.
\end{aligned}
\end{equation}
The first two commuting Hamiltonians of the Toda system are coming from the leading two non-trivial coefficients:
\begin{subequations}
\begin{align}
h_1&=\sum_\omega \left( \epsilon-a_{\omega+1}+\epsilon\nu_{\omega} \right), \\
h_2&=
\sum_{\omega}\left[\frac{1}{2}(\epsilon-a_{\omega+1}+\epsilon\nu_{\omega})^2-\frac{1}{2}(a_{\omega}-\epsilon)^2+\epsilon D_\omega+\mathfrak{q}_\omega\Lambda^2\right]  \nonumber \\
&=\sum_{\omega}\left[\frac{1}{2}(\epsilon-a_{\omega+1}+\epsilon\nu_{\omega})^2+\epsilon D_\omega+\mathfrak{q}_\omega\right]-\frac{1}{2}\sum_{\alpha=1}^Np_\alpha^2.
\end{align}
\end{subequations}
Now using the definitions in \eqref{kv-data}, we see that if we treat $h_1$ as an operator, when acting on orbifolded instanton partition function and using \eqref{split q}, we may replace 
\begin{align}
    \nu_\omega=k_{\omega}-k_{\omega+1}\to\mathfrak{q}_\omega\frac{\partial}{\partial\mathfrak{q}_{\omega}}-\mathfrak{q}_{\omega+1}\frac{\partial}{\partial\mathfrak{q}_{\omega+1}}.
\end{align}
Let us define
\begin{equation}\label{q-x}
\mathfrak{q}_\omega=\Lambda^2e^{\rx_{\omega+1}-\rx_{\omega}}.
\end{equation}
and we may replace
\begin{align}\label{nu to partial}
    \nu_\omega\to\frac{\partial}{\partial\rx_{\omega+1}}
\end{align}
\paragraph{}
As for $h_2$, by the definition of $D_\omega$, we can set $\sum_\omega D_{\omega}=0$. 
Now we have
\begin{equation}
h_2=\frac{\epsilon^2}{2}\sum_{\alpha=1}^{N} \left(\frac{\partial}{\partial \rx_{\alpha}}+1-\frac{a_{\omega}}{\epsilon}\right)^2+\Lambda^2\left(\sum_{\omega=\alpha}^{N} e^{\rx_{\alpha+1}-\rx_{\alpha}}\right);\quad \rx_{N+\alpha}=\rx_\alpha,
\end{equation}
acting on orbifolded instanton partition function.
If we further take the classical limit $\epsilon\to0$, the kinetic term becomes $\sum_{\alpha}\frac{1}{2}a_{\alpha}^2$. This means $a_\alpha$ must be real in order to have a non-negative kinetic energy term. 
We have thus recovered the Hamiltonians of the periodic Toda chain eq.~\eqref{H-Toda}. $T(x)$ defined in eq.~\eqref{Tfunc} using the instanton partition function is indeed the characteristic polynomial of the corresponding integrable system. Calculation without taking NS-limit can be found in \cite{Jeong:2017pai}.

\section{The characteristic polynomial of eCM Model}\label{X-function}
\subsection{The need for $\mathbb{X}$-function}
\paragraph{}
With success of $\widehat{A}_{N-1}$ Toda system, we would like to ask whether $T(x)$ defined for $\mathcal{N}=2^*$ system in \eqref{firstT} can similarly reproduce commuting Hamiltonians of the eCM system, hence be identified as the characteristic polynomial? 
\paragraph{}
The short answer is: NO. This is because $T(x)$ defined earlier in \eqref{firstT} is not a finite degree polynomial. There exist additional poles coming from $Y(x-m)Y(x+m+\epsilon)$ in the numerator, 
which render it non-analytic.
Explicit calculation also verifies our claim.  As with Toda lattice, here we introduce the full surface defect on $\mathbb{C}_1\subset \mathbb{C}_1 \times \mathbb{C}_2 = \mathbb{R}^4 $ and $\mathbb{Z}_N$ orbifolding which maps $(\mathbf{z}_1,\mathbf{z}_2)\to(\mathbf{z}_1,\zeta \mathbf{z}_2)$, $\zeta^N=1$. Following similar orbifolding procedures from eq.~\eqref{Y_w} to eq.~\eqref{T-x-large}, we found for $\omega =0, \dots N-1$:
\begin{equation}
\begin{aligned}
	T_\omega(x)
	=&{Y_{\omega+1}(x+\epsilon)}+\mathfrak{q}_\omega\frac{Y_\omega(x-m)Y_{\omega+1}(x+m+\epsilon)}{Y_\omega(x)} \\
	=& \left[x+\epsilon-a_{\omega+1}+\mathfrak{q}_\omega\left((x+m+\epsilon-a_{\omega+1})\frac{x-a_\omega-m}{x-a_\omega}\right)\right]\exp\left(\frac{\epsilon}{x}\nu_\omega+\frac{\epsilon}{x^2}D_\omega+\cdots\right) \\
	=& (1+\mathfrak{q}_\omega)(x+\epsilon-a_{\omega+1})\exp\left(\frac{\epsilon}{x}\nu_\omega+\frac{\epsilon}{x^2}D_\omega+\cdots\right)+\frac{\mathfrak{q}_\omega}{x}(-m(m+\epsilon)+ma_{\omega+1}-a_\omega) \\
	=&(1+\mathfrak{q}_\omega)\left[x+\epsilon-a_{\omega+1}+\epsilon\nu_{\omega}+\frac{1}{x}\left[\frac{1}{2}\epsilon^2\nu_{\omega}^2-\epsilon\nu_{\omega}a_{\omega+1}+\epsilon_1D_\omega-\epsilon^2\nu_\omega \right]\right]+\frac{\mathfrak{q}_\omega}{x}(-m(m+\epsilon)+ma_{\omega+1}-a_\omega)+\cdots
\end{aligned}
\end{equation}
If we normalize $T_{\omega}(x)$ by  $1+\mathfrak{q}_\omega$ to set the coefficient of the leading term to unity:
\begin{equation}\label{first T_w}
	\prod_{\omega}\frac{T_\omega(x)}{1+\mathfrak{q}_\omega}=x^N+h_1x^{N-1}+h_2x^{N-2}+\cdots
\end{equation}
then we will have the first few $h_j$'s as
\begin{subequations}
\begin{align}
	h_1&=-\sum_\omega (a_{\omega}-\epsilon), \\
	h_2 
	&=\frac{1}{2}h_1^2-\frac{1}{2} \sum_{\omega}(a_{\omega}-\epsilon)^2+\epsilon_1D_\omega+\frac{\mathfrak{q}_\omega}{1+\mathfrak{q}_\omega}(-m(m+\epsilon)+ma_{\omega+1}-a_\omega). 
\end{align}
\end{subequations}
Here we see that $h_2$ obtained clearly does not resemble eq.~\eqref{H-CS}. Also since eq.~\eqref{firstT} consists poles, eq.~\eqref{first T_w} is NOT a polynomial.
\paragraph{}
The failure of reproducing the correct eCM Hamiltonian leads us to consider a certain modification of $T(x)$,  which we will denote it as $\mathbb{X}(x)$.%
\footnote{%
The $\bbX$-function with generic $(\epsilon_1,\epsilon_2)$ and $\vec{\lambda}$ is also known as the fundamental $qq$-character of $\widehat{A}_0$ quiver~\cite{Nikita:I,Kimura:2015rgi}, which is reduced to the corresponding $q$-character in the NS limit, $\epsilon_2 \to 0$~\cite{Nikita-Pestun-Shatashvili}.}
In the construction of this function, we again temporarily restore the $\epsilon_2$ dependence:
\begin{equation}\label{Xbb-obs}
\mathbb{X}(x)={Y(x+\epsilon_+)}\sum_{\{{\mu}\}}\mathfrak{q}^{|{\mu}|}{B}[{\mu}]\prod_{(\bi,\bj)\in\mu}\frac{Y(x+s_{\bi\bj}-m)Y(x+s_{\bi\bj}+m+\epsilon_+)}{Y(x+s_{\bi\bj})Y(x+s_{\bi\bj}+\epsilon_+)}
\end{equation}
where $\epsilon_+=\epsilon_1+\epsilon_2$, and ${\mu}$ is one single Young diagram, we denote it as:
\begin{equation}
{\mu}=(\mu_1,\mu_2,\dots,\mu_{\ell(\mu)}).
\end{equation}
Since $\mu$ is only one Young diagram, we will not use vector notation like $\vec{\lambda} = (\lambda^{(1)},\ldots,\lambda^{(N)})$, the latter denotes a vector with $N$ Young diagrams $\lambda^{(\alpha)}$, $\alpha=1,\dots,N$ as its entries. Note that ${\mu}$ has no relation to $\vec{\lambda}$ labeling the fixed point on the original instanton moduli space. We will see later in chapter \ref{eDCM-Origami} that Young diagram $\mu$ has an interpretation as the ``dual'' instanton configuration in the eight-dimensional gauge origami construction. Each box of $\mu$ is labeled by:
\begin{equation}
s_{\bi\bj}=(\bi-1)m-(\bj-1)(m+\epsilon_+)
\end{equation}
where $\bi=1,\dots,\ell(\mu)$ and $\bj=1,\dots,\mu_\bi$ with a given $\bi$. Let us define
\begin{equation}
{B}[{\mu}]=\prod_{({\bi,\bj})\in\mu}B_{1,2}(mh_{\bi\bj}+\epsilon_+ a_{\bi\bj});\qquad B_{12}(x)=1+\frac{\epsilon_1\epsilon_2}{x(x+\epsilon_+)}.
\end{equation}
Here $a_{\bi\bj}=\mu_{\bi}-\bj$ denotes the ``arm'' associated with a given a box $(\bi,\bj)$ in a Young diagram, $l_{\bi\bj}=\mu_{\bi\bj}^T-\bi$ for the ``leg'' associated with the same box. We have also defined $h_{\bi\bj}=a_{\bi\bj}+l_{\bi\bj}+1$. Under NS limit, $\lim_{\epsilon_2\to0}B[{\mu}]=1$ for all $\mu$. 
The relation between $T(x)$ defined in \eqref{firstT} and $\mathbb{X}(x)$ defined in eq.~\eqref{Xbb-obs} has been mentioned in \cite{FJMM}.
One may identify our $T(x)$ with the $T$-function denoted by $T_{6v}$ there and $\mathbb{X}(x)$ with the other $T$-function denoted as $\mathscr{T}$. Comparing with \cite{Nikita:V} and \cite{Nikita:I}  , we see that $\mathbb{X}$-function \eqref{Xbb-obs} is the qq-character of $\mathcal{N}=2^*$ system defined on limit shape, and becomes the q-character when NS limit is taken.

\paragraph{}
We will now continue to take NS-limit and show that $\mathbb{X}(x)$ is a degree $N$ polynomial.
Using the large $x$-asymptotic behavior of $Y(x)$, it is obvious that $\mathbb{X}(x)$ is of order $N$. 
To prove its analyticity, let us consider one specific $\mu$ configuration $\mu=(\mu_1,\mu_2,\cdots,\mu_{\ell(\mu)})$ under NS-limit, its contribution to $\bbX(x)$ is denoted as
\begin{equation}
\begin{aligned}
\mathbb{X}(x)[\mu]&=\mathfrak{q}^{|\mu|}Y(x+\epsilon)\prod_{\bi=1}^{\ell(\mu)}\prod_{\bj=1}^{\mu_\bi}\frac{Y(x+s_{\bi\bj}-m)Y(x+s_{\bi\bj}+m+\epsilon)}{Y(x+s_{\bi\bj})Y(x+s_{\bi\bj}+\epsilon)} \\
&=\mathfrak{q}^{|\mu|}Y(x+{\ell(\mu)}m)\prod_{\bi=1}^{\ell(\mu)}\frac{Y(x+(\bi-1)m-\mu_\bi(m+\epsilon)+\epsilon)}{Y(x+\bi m-\mu_\bi(m+\epsilon)+\epsilon)} \\
&=\mathfrak{q}^{|\mu|}\frac{Q(x+{\ell(\mu)}m)}{Q(x+{\ell(\mu)}m-\epsilon)}\prod_{\bi=1}^{\ell(\mu)}\frac{Q(x+(\bi-1)m-\mu_\bi(m+\epsilon)+\epsilon)}{Q(x+(\bi-1)m-\mu_\bi(m+\epsilon))}\frac{Q(x+\bi m-\mu_\bi(m+\epsilon))}{Q(x+\bi m-\mu_\bi(m+\epsilon)+\epsilon)},
\end{aligned}
\end{equation}
such that the total $\mathbb{X}$-function is given by:
\begin{align}\label{sum-X}
 \mathbb{X}(x) = \sum_{\{\mu \}} \mathbb{X}(x)[\mu].
\end{align}
The poles of $\mathbb{X}(x)[\mu]$ are located at 
\begin{itemize}
	\item $\{x_{\alpha i}-{\ell(\mu)}m+\epsilon\}$ from zeros of $Q(x+{\ell(\mu)}m-\epsilon)$,
	\item $\{x_{\alpha i}-(\bi-1)m+\mu_\bi(m+\epsilon)\}$ from zeros of $Q(x+(\bi-1)m-\mu_\bi(m+\epsilon))$, and
	\item $\{x_{\alpha i}-\bi m+\mu_\bi(m+\epsilon)-\epsilon\}$ form zeros of $Q(x+\bi m-\mu_\bi(m+\epsilon))$,
\end{itemize}
where $x_{\alpha i}$ is defined in \eqref{Def:xx0}. For each $\bi$ there exists an infinity number of poles from infinity many $\{x_{\alpha i}\}$, $\alpha=1,\dots,N$, $i\in\mathbb{N}$.
\paragraph{}
Let us focus on the poles located at $x_{\alpha i}-(\bl-1)m+\mu_\bl(m+\epsilon)$ of some $1\leq\bl\leq \ell(\mu)$. 
Adding an additional box to $\mu$ located at $(\bl,\mu_\bl)$ gives a new Young diagram $\mu'=(\mu_1,\cdots,\mu_{\bl-1},\mu_\bl+1,\mu_{\bl+1},\cdots,\mu_{\ell(\mu)})$, whose contribution to $\mathbb{X}(x)$ is:
\begin{equation}
\begin{aligned}
\mathbb{X}(x)[\mu']=\mathbb{X}(x)[\mu] \times\mathfrak{q}\frac{Y(x+(\bl-2)m-\mu_\bl(m+\epsilon))Y(x+(\bl-1)m-(\mu_\bl-1)(m+\epsilon))}{Y(x+(\bl-1)m-\mu_\bl(m+\epsilon))Y(x+(\bl-1)m-\mu_\bl(m+\epsilon)+\epsilon_1)}.
\end{aligned}
\end{equation}
Both $\mathbb{X}(x)[\mu']$ and $\mathbb{X}(x)[\mu]$ are contained in $\bbX(x)$ \eqref{sum-X} and share the same poles $x_{\alpha i}-(\bl-1)m+\mu_\bl(m+\epsilon)$. The sum of the two Young diagram contributions gives us
\begin{align}
\mathbb{X}(x)[\mu]+\mathbb{X}(x)[\mu'] 
& =\mathbb{X}(x)[\mu]\left[1+\mathfrak{q}\frac{Y(x+(\bl-2)m-\mu_\bl(m+\epsilon))Y(x+(\bl-1)m-(\mu_\bl-1)(m+\epsilon))}{Y(x+(\bl-1)m-\mu_\bl(m+\epsilon))Y(x+(\bl-1)m-\mu_\bl(m+\epsilon)+\epsilon_1)}\right] 
\nonumber \\
& \longrightarrow \ 0 \qquad \left(x \to x_{\alpha i}-(\bl-1)m+\mu_\bl(m+\epsilon)\right).
\end{align}
The poles located at $x_{\alpha i}-(\bl-1)m+\mu_\bl(m+\epsilon)$ from $\mathbb{X}(x)[\mu]$ are now canceled by the denominator using eq.~\eqref{A0T}. The other two sets of poles can be dealt with similarly. Since $\mathbb{X}(x)$ is summed over all Young diagram configuration, $\mathbb{X}(x)$ is analytic.

\subsection{Commuting Hamiltonians from $\mathbb{X}(x)$}
\paragraph{}
Finally we would like to see that the correct commuting Hamiltonians of the eCM system can be obtained directly from $\mathbb{X}(x)$ we just constructed.
Following the same procedure we performed with Toda system in the end of previous section, a full-type surface defect is again introduced in $\mathbb{C}_{1} \subset \bR^4$ with orbifolding.
Each orbifolded copy of $\bbX(x)$ under NS-limit becomes:
\begin{equation}\label{X_w}
\mathbb{X}_\omega(x)={Y_{\omega+1}(x+\epsilon)}\sum_{\{\mu\}}\mathbb{B}^{\mu}_\omega\prod_{(\bi,\bj)\in\mu}\frac{Y_{\omega+1-\bj}(x+s_{\bi\bj}-m)Y_{\omega+1-\bj+1}(x+s_{\bi\bj}+m+\epsilon)}{Y_{\omega+1-\bj}(x+s_{\bi\bj})Y_{\omega+1-\bj+1}(x+s_{\bi\bj}+\epsilon)}.
\end{equation}
The full $\mathbb{X}$-function can be recovered via
\begin{equation}\label{X-X_w}
	\mathbb{X}(x)=\prod_\omega\mathbb{X}_\omega(x),
\end{equation}
where each $\mathbb{X}_\omega(x)$ is of degree one.
\paragraph{}
Here we would like to address further the factor $\mathbb{B}_\omega^{{\mu}}$ appearing in the summation, it is the orbifolded version of  $\mathfrak{q}^{|\mu|}B[\mu]$ appearing in \eqref{Xbb-obs}. Consider the summation over all possible partition configurations, which we denote as ${\bf{B}}$:
\begin{equation}\label{B}
\mathbf{B}=\sum_{\{\mu\}}\mathfrak{q}^{|\mu|}{B}[\mu]
\end{equation}
This is equivalent to a single $\mathcal{N}=2^*$ $U(1)$ instanton partition function with $(m,-m-\epsilon_+)$ identified as its $\Omega$-background parameters as pointed out in \cite{Nikita:I}.
This observation will eventually lead us to so called ``Gauge Origami'' construction, for which we will discuss in Section~\ref{GaugeOrigami}. 
After orbifolding, each individual $\mathbb{B}_\omega^{\mu}$ becomes
\begin{equation}\label{B_w}
\mathbb{B}^{\mu}_\omega=\left.\prod_{(\bi,\bj)\in\mu}\mathfrak{q}_{\omega+1-\bj}B_1(mh_{\bi\bj} )\right|_{a_{\bi\bj}=0}
\end{equation}
with 
\begin{equation}
B_1(x)=1+\frac{\epsilon}{x},
\end{equation}
and here we define the following:
\begin{equation}\label{K_w}
\begin{aligned}
&K^\mu_\omega:=\{(\alpha,(\bi,\bj)) \mid \alpha=1,\dots,N;\quad(\bi,\bj)\in\mu;\quad\alpha-\bj+1\equiv\omega \text{ mod}(N)\}; \\
& k^\mu_\omega=|K^\mu_\omega|;\quad\nu^\mu_\omega=k^\mu_{\omega}-k^\mu_{\omega+1}.
\end{aligned}
\end{equation}
Let us define a new set of variables
(instead of \eqref{q-x})
\begin{align}\label{q-z}
    \mathfrak{q}_\omega=\frac{z_\omega}{z_{\omega-1}};\quad z_{\omega+N}=\mathfrak{q}z_{\omega}
\end{align}
such that \eqref{B_w} can  now be rewritten as
\begin{align}
    \mathbb{B}^{\mu}_\omega(\vec{z};\tau)=\prod_{l=1}^{\mu_1}\prod_{h=1}^{\mu^T_{\bl}-\mu^T_{l+1}}\frac{z_\omega}{z_{\omega-l}}B_1(mh).
\end{align}
One way to think about this configuration is that the orbifolding now splits the instanton partition into $N$ copies of $U(1)$ sub-partitions. Each element in $K_\omega$ is counted by orbifolded coupling $\mathfrak{q}_\omega$ instead of the original $\mathfrak{q}$. To evaluate the summation over all possible Young diagrams, we will introduce a new representation for a Young diagram $\mu$:
\begin{equation}
    \mu=(1^{l_0}2^{l_1}\dots (N-1)^{l_{N-2}}(N)^l).
\end{equation}
Each $l_{r-1}=\sum_{J=0}^\infty l_{r-1,J}$, where $l_{r-1,J}=\left(\mu_{r+NJ}^{T}-\mu^T_{r+1+NJ}\right)$ is the difference between number of boxes of two neighboring columns, $r = 1,\ldots, N-1$ and the last one $\bl=\sum_{J=1}^\infty\mu_{NJ}^T$ counts for how many times a full combination of $\mathfrak{q}_0\cdots\mathfrak{q}_{N-1}=\mathfrak{q}$ shows up.
Define the summation over all possible partition configuration of each $\omega$ as:
\begin{equation}\label{B-form}
\begin{aligned}
\mathbb{B}_\omega(\vec{z};\tau)
=\sum_{\{\mu\}}\mathbb{B}_\omega^{\mu}(\vec{z};\tau)
=\sum_{l_0,\dots,l_{N-1},\bl\geq0}\prod_{\alpha=0}^{N-1}\frac{\left(l_\alpha+\frac{\epsilon_1}{m}\right)!}{(l_\alpha)!\left(\frac{\epsilon_1}{m}\right)!}\left(\frac{z_\omega}{z_\alpha}\right)^{l_\alpha}\mathfrak{q}^\bl,
\end{aligned}
\end{equation}
and their total product
\begin{equation}\label{B=QF}
\mathbb{B}(\vec{z}, \tau)=\prod_\omega\mathbb{B}_\omega (\vec{z}, \tau)=\mathbb{Q}^{\frac{m+\epsilon}{m}}(\vec{z};\tau)F(\tau),
\end{equation}
is the orbifolded version of ${\bf{B}}$ defined in eq.~\eqref{B}, i.e. the orbifolded instanton partition function of $U(1)$ $\mathcal{N}=2^*$ theory in the NS limit. The function $\mathbb{Q}(\vec{z};\tau)$ is defined in \eqref{Q} in terms of elliptic theta functions. The explicit form of the $\vec{z}$ independent function $F(\tau)$ will not be used in the following derivation of commuting Hamiltonians and we will show it can be absorbed by shifting the zero point energy.
\paragraph{}
We again consider the large $x$ expansion of $\mathbb{X}_\omega$ in \eqref{X_w} and we normalize $\mathbb{X}_\omega(x)$ with respect to the coefficient of the leading $x$ term, which is $\mathbb{B}_\omega(\vec{z}, \tau)$.
A similar computation yields:  
\begin{equation}
\begin{aligned}
	&\frac{1}{\mathbb{B}_\omega(\vec{z}, \tau)}\mathbb{X}_\omega(x)
	=x+\epsilon-a_{\omega+1}+\epsilon_1\nu_\omega+ \\
	&\frac{1}{x}\left[\frac{1}{2}(\epsilon\nu_\omega-a_{\omega+1})^2-\frac{1}{2}(a_{\omega+1})^2+\epsilon D_\omega-m\sum_{\{\mu\}}\frac{\mathbb{B}_{\omega}^{\mu}(\vec{z},\tau)}{\mathbb{B}_\omega(\vec{z},\tau)}\sum_{\omega'=0}^{N-1}\left((m+\epsilon)k^\mu_{\omega'}+\left(\epsilon\nu_{\omega'}-a_{\omega'+1}\right)\nu^\mu_{\omega'}\right)\right]+\dots
\end{aligned}
\end{equation}
As stated in \eqref{X-X_w}, the full $\mathbb{X}$-function carrying the information of the conserved Hamiltonians is the product of all the orbifolded pieces $\mathbb{X}_\omega(x)$. 
From above we will take the normalization to be:
\begin{equation}
	\frac{\mathbb{X}(x)}{\mathbb{B}(\vec{z}, \tau)}=\frac{ \prod_{\omega} \mathbb{X}_\omega(x)}{\prod_{\omega}{\mathbb{B}_\omega}(\vec{z},\tau)}=x^N+h_1x^{N-1}+h_2x^{N-2}+\cdots+h_N.
\end{equation}
To express the commuting Hamiltonians, let us define the following derivative operators for $\omega=1,\dots,N$:
\begin{equation}
\nabla^\mathfrak{q}_\omega=\mathfrak{q}_\omega\frac{\partial}{\partial \mathfrak{q}_\omega}.
\end{equation}
and differential operators for $z$:
\begin{equation}\label{Laplacian}
\nabla_{\omega}^z=z_\omega\frac{\partial}{\partial z_\omega};\quad
\Delta_{\vec{z}}=\sum_{\omega=0}^{N-1}\nabla_{\omega}^z\nabla_{\omega}^z.
\end{equation}
Based on \eqref{q-z}, it implies the relation:
\begin{equation}
\nabla_{\omega}^z=\nabla^\mathfrak{q}_\omega-\nabla^\mathfrak{q}_{\omega+1}.
\end{equation}
\paragraph{}
Using definition of $\mathbb{B}_\omega (\vec{z}, \tau)$ in \eqref{B_w}, and \eqref{K_w}, we can express the first  commuting Hamiltonians as:
\begin{subequations}
\begin{align}
	h_1&=\sum_{\omega=0}^{N-1} (\epsilon-a_{\omega+1}+\epsilon\nu_{\omega})=N\epsilon+\sum_{\omega=0}^{N-1}P_\omega; \\
	h_2&=\frac{1}{2}h_1^2+\sum_{\omega=0}^{N-1}-\frac{1}{2}(a_{\omega})^2-m\left((m+\epsilon)\sum_{\omega'=0}^{N-1}\nabla_{\omega'}^{\mathfrak{q}}+\sum_{\omega'=0}^{N-1}\left(\epsilon\nu_{\omega'}-a_{\omega'+1}\right)\nabla_{\omega'}^z\right)\log\mathbb{B}(\vec{z};\tau), \nonumber\\
	&=\frac{1}{2}\sum_{\omega=0}^{N-1}P_\omega^2 -\frac{1}{2}\sum_{\omega=0}^{N-1}(a_{\omega})^2-m\left((m+\epsilon)\sum_{\omega=0}^{N-1}\nabla_{\omega}^{\mathfrak{q}}+\sum_{\omega=0}^{N-1}P_{\omega}\nabla_{\omega}^z\right)\log\mathbb{B}(\vec{z};\tau).
\end{align}
\end{subequations}
Again like the case of Toda, we may treat $h_1$ as operator acting on orbifolded instanton partition function and replace $\nu_\omega\to\nabla^z_\omega$, thus we have the momentum.
\begin{align}
    P_\omega=\epsilon\nabla^z_\omega-a_{\omega+1}.
\end{align}
We claim that we have recovered eq.~\eqref{H-CS} up to the following canonical transformation between the generalized coordinate $q$ and its conjugate momentum $P$, which satisfy the commutation relation $[q,P]=\epsilon$: 
\begin{align}\label{canonical trans}
	H 
    & = \frac{1}{2}P^2+Pf(q)+V(q)	\nonumber \\
    & =\frac{1}{2}\left(P+f(q)\right)^2+V(q)+\frac{[P,q]}{2}f'(q)-\frac{1}{2}f(q)^2 \nonumber \\
    & = \frac{1}{2}(P+f(q))^2+V(q)-\frac{\epsilon}{2}f'(q)-\frac{1}{2}f(q)^2.
\end{align}
We can rewrite potential terms in $h_2$ as 
\begin{equation}\label{CM Potential}
	V(\vec{z})=-m(m+\epsilon)\sum_{\omega=0}^{N-1}\nabla_{\omega}^{\mathfrak{q}}\log\mathbb{B}(\vec{z};\tau)-\frac{1}{2}m^2\sum_{\omega=0}^{N-1}\left(\nabla_{\omega}^z\log\mathbb{B}(\vec{z};\tau)\right)^2-\frac{1}{2}m\epsilon\Delta_{\vec{z}}\log\mathbb{B}(\vec{z};\tau).
\end{equation}
By using eq.~\eqref{B=QF} and eq.~\eqref{Heat eq for Q}, we may finally rewrite
\begin{equation}
\begin{aligned}
	h_2
	&=\sum_{\omega=0}^{N-1}\frac{P_{\omega+1}^2}{2}+\frac{(m+\epsilon)^2-\epsilon(m+\epsilon)}{2}\Delta_{\vec{z}}\log\mathbb{Q}(\vec{z};\tau)- Nm(m+\epsilon_1)\nabla^\mathfrak{q}F(\tau) \\
	&=\frac{1}{2}\sum_{\alpha=1}^N{P_\alpha^2}+m(m+\epsilon)\sum_{\alpha>\beta}\wp(z_\alpha/z_\beta;\tau)-Nm(m+\epsilon_1)\nabla^\mathfrak{q}F(\tau),
\end{aligned}
\end{equation}
in particular $F(\tau)$ may be removed by shifting the zero energy level, its explicit form is not important as noted earlier.
We have thus successfully recovered the quantum eCM Hamiltonian given in eq.~\eqref{H-CS}. 
Here we summarize the explicit parameter identifications in $\cN=2^*$ $SU(N)$ gauge theory and the $N$-particle eCM system:
\begin{center}
\begin{tabular}{|c|c|c|}
	\hline
	& Gauge Theory & Integrable System \\ \hline\hline
	$a_\alpha$ & Coulomb Moduli & Momenta \\ \hline
	$\tau$ & Complex gauge coupling & Elliptic modulus  \\ \hline
	$\epsilon$ & $\Omega$-deformation parameter & Planck constant \\ \hline
	$m$ & Adjoint mass & Coupling constant \\ \hline
	$N$ & Gauge group rank & Number of particles \\ \hline
	$z_\alpha$ & Ratio between orbifolded couplings & Exponentiated coordinates  \\
	\hline
\end{tabular}
\end{center}
By using second property in \eqref{q-z} that $z_{\omega+N}=\mathfrak{q}z_\omega$, the coordinates $\{z_\alpha\}$ and complex coupling $\mathfrak{q}=e^{2\pi i\tau}$ are independent. 
\paragraph{}
Let us end this section by commenting that one way to identify eigenfunction of $h_2$ is to use the fact $\mathbb{X}(x)$ is $\mathcal{N}=2^*$ q-character. In the NS-limit, the VEV of q-character is dominated by following limiting shape configuration
\begin{align}
    t(x)=\langle\mathcal{X}(x)\rangle=\frac{\sum_{\vec{\lambda}}\mathcal{X}(x)[\vec{\lambda}]\mathcal{Z}_\text{inst}[\vec{\lambda}]}{\mathcal{Z}_\text{inst}}
    =\frac{\mathbb{X}(x)\mathcal{Z}_\text{inst}[\vec{\lambda}_*]}{\mathcal{Z}_\text{inst}[\vec{\lambda}_*]}
\end{align}
where $t(x)=x^N+E_1x^{N-1}+E_2x^{N-2}+\cdots+E_N$.
When treating Hamiltonians as the operators (and thus $\mathbb{X}(x)$), we have
\begin{align}
    \mathbb{X}(x)\mathcal{Z}_\text{inst}[\vec{\lambda}_*](\vec{\rx})=t(x)\mathcal{Z}_\text{inst}[\vec{\lambda}_*](\vec{\rx}).
\end{align}
By matching the coefficients, we conclude $\mathcal{Z}_\text{inst}[\vec{\lambda}_*]$ is the eigenfunction of Hamiltonians
\begin{align}
    h_i\mathcal{Z}_\text{inst}[\vec{\lambda}_*](\vec{\rx})=E_i\mathcal{Z}_\text{inst}[\vec{\lambda}_*](\vec{\rx});\quad i=1,2,\dots,N.
\end{align}
The canonical transformation performed in \eqref{canonical trans} gives an additional factor to the orbifolded instanton partition function. $h_2$ has the eigenfunction as:
\begin{align}
    \Psi(\vec{\rx})=\mathbb{Q}^{-\frac{m+\epsilon}{\epsilon}}(\vec{\rx})\mathcal{Z}_\text{inst}[\vec{\lambda}_*](\vec{\rx});\quad h_2\Psi(\vec{\rx})=E_2\Psi(\vec{\rx}).
\end{align}
Detailed calculations and discussion can be found in \cite{Nikita-Shatashvili,Nikita:V}.

\section{Elliptic Double Calogero-Moser System from Gauge Origami}\label{GaugeOrigami}
\paragraph{}
Let us begin by introducing the basic information about the elliptic double Calogero-Moser system (edCM),
it is an one dimensional quantum mechanical system consisting of $P=N+M$ particles  governed by the following Hamiltonian:
\begin{equation}\label{DCM}
\begin{aligned}
    \frac{1}{\hbar^2}\hat{H}_\text{edCM}=
    & -\frac{1}{2}\sum_{\alpha=1}^N\frac{\partial^2}{\partial \rx_\alpha^2}-\frac{k}{2}\sum_{\beta=1}^M\frac{\partial^2}{\partial \ry_\beta^2} \\
    &+k(k+1)\sum_{1\leq \alpha'<\alpha\leq N}\wp(\rx_\alpha-\rx_{\alpha'}) +\left(\frac{1}{k}+1\right)\sum_{1\leq \beta'<\beta\leq M}\wp(\ry_\beta-\ry_{\beta'})+(k+1)\sum_{\alpha=1}^N\sum_{\beta=1}^M\wp(\rx_\alpha-\ry_\beta).
\end{aligned}
\end{equation}
The constant $k$ is the ratio of masses between two sets of identical particles, {i.e. the mass of the first $N$-particles labeled by $\{\rx_\alpha\}_{\alpha=1}^N$ is $k$ times of the mass of the remaining $M$-particles labeled by $\{\ry_{\beta}\}_{\beta=1}^M$. }Simultaneously $k$ also acts as a single coupling constant. The Hamiltonian~\eqref{DCM} was initially mentioned in the context of the gauge origami in~\cite{Nikita:V}, and the trigonometrical limit of eq.~\eqref{DCM} is studied in various papers such as \cite{SV,HiJack}.
Notice that \eqref{DCM} inherits the following symmetry: Swapping $\{\rx_\alpha\}\leftrightarrow\{\ry_\beta\}$ while simultaneously flipping $k\leftrightarrow\frac{1}{k}$ (up to over all $k$ factor).
\paragraph{}
Let us look more closely at Hamiltonian given in eq.~\eqref{DCM}. In particular comparing with eq.~\eqref{H-CS} and coefficients of their potential when there is only one group of particles. 
\begin{itemize}
	\item As $M=0$, we identify $k=\frac{m}{\hbar}$;
	\item As $N=0$, we identify $\frac{1}{k}=\frac{m}{\hbar}$.
\end{itemize}
Here we see that the meaning of ``classical'' limit is somewhat ambiguous among the two sets of particles. For particles labeled by $\{\rx_\alpha\}_{\alpha=1}^N$, the classical limit means taking $k\gg1$ while keeping $m$ finite. As for particles labeled by $\{\ry_{\beta}\}_{\beta=1}^M$, the classical limit is taken under $k\ll1$. This is the first hint that Hamiltonian in eq.~\eqref{DCM} has no natural classical limit. Suppose we take 
$k=\frac{m}{\hbar}$,
taking the classical limit $\hbar\to0$ is equivalent to have $k\gg1$. 
In such a limit, the mass of the first $N$ particles labeled by $\{\rx_\alpha\}_{\alpha=1}^N$ is much heavier than the remaining $M$ particles. In the classical approach, those objects with much larger mass can be treated as non-dynamical in the leading order. 
Eq.~\eqref{DCM} now becomes:
\begin{equation}
	\hat{H}_\text{edCM} \ \stackrel{k \to \infty}{\longrightarrow} \ -\frac{k}{2}\sum_{\beta=1}^M\frac{\partial^2}{\partial \ry_\beta^2}+k\sum_{\alpha=1}^N\sum_{\beta=1}^M\wp(\rx_\alpha-\ry_\beta)+k^2\sum_{1\leq \alpha'<\alpha\leq N}\wp(\rx_\alpha-\rx_{\alpha'}).
\end{equation}
Even though the last term is of $k^2$ order, it is just a constant as the heavy particles are non-dynamical. The resultant quantum Hamiltonian describes $M$ non-interacting particles in a potential well. Similar argument applies to $k=\frac{\hbar}{m}\ll1$. We conclude that the system defined by eq.~\eqref{DCM} has no classical limit. In particular taking large mass limit with $\hbar$ fixed is equivalent to take large $k$. Thus unlike eCM we do not recover double Toda under such limit by the fact edCM has no classical limit.
This analysis also indicates that the connection of such an inherently quantum system with the supersymmetric gauge theories is much more subtle as we will reveal shortly.

\subsection{Quantum Integrability of Elliptic Double Calogero-Moser System}
\paragraph{}
Before constructing the supersymmetric gauge theory associated with the edCM system however, let us first further investigate its integrability and we will employ the so-called  Dunkl operators \cite{Dunkl:}. 
The Dunkl operators are quantum version of Lax pairs \cite{DP2,Pasquier,EllipticDunkl} which pairwise commute. In particular the Dunkl operators for Calogero-Sutherland integrable models were explicitly worked out in \cite{EllipticDunkl}, and their equivalence to the quantum pair Lax operators \cite{Bordner:1999xq} was shown in \cite{Khastgir:2000ig} for all root systems. To explicitly define them, let us consider the following family of functions:
\begin{equation}
 \sigma_t(x) =
  \frac{\theta_{11}(x-t)\theta_{11}'(0)}{\theta_{11}(x)\theta_{11}(-t)} ;
  \quad t\in\mathbb{C}/(\mathbb{Z}\oplus\tau\mathbb{Z}),
\end{equation}
where $\tau$ is a modular parameter and $\theta_{11}$ is the theta function defined in \eqref{theta2} (Recall that we have identify complex gauge coupling with  elliptic modulus at the end of previous chapter.).
The function $\sigma_t(x)$ has the following properties
\begin{subequations}\label{sigma properties}
	\begin{align}
    &\sigma_t(x+2\pi i)=\sigma_t(x), \\
    &\sigma_t(x)=-\sigma_{-t}(-x)\label{sigma minus}, \\
    &\sigma_t(x)=-\sigma_x(t) ,\\
    &\sigma_t(x)\sigma_{-t}(x)=\wp(x)-\wp(t), \\
    &\lim_{t\to0}\frac{d}{dx}\sigma_t(x)=-\wp(x)-2\zeta\left(\frac{1}{2}\right).
    \end{align}
\end{subequations}
Let $t_\alpha$, $\alpha=1,\dots,N$,  and $u_\beta$, $\beta=1,\dots,M$, be $P=N+M$ complex numbers,
$t_\alpha,\,u_\beta\in\mathbb{C}/(\mathbb{Z}\oplus\tau\mathbb{Z})$. 
The elliptic double Dunkl operators are defined as:
\begin{subequations}\label{Dunkl}
\begin{align}
d^\rx_\alpha
&=\frac{\partial}{\partial \rx_\alpha}+k\sum_{\substack{\alpha'=1 \\ (\alpha'\neq \alpha)}}^N\sigma_{t_\alpha-t_{\alpha'}}(\rx_\alpha-\rx_{\alpha'})S^{\rx\rx}_{\alpha\alpha'}+\sum_{\beta=1}^{M}\sigma_{t_\alpha-u_\beta}(\rx_\alpha-\ry_\beta)S^{\rx\ry}_{\alpha\beta}, \\
d^\ry_\beta
&=k\frac{\partial}{\partial \ry_\beta}+k\sum_{\alpha=1}^N\sigma_{u_\beta-t_\alpha}(\ry_\beta-\rx_\alpha)S_{\alpha\beta}^{\rx\ry}+\sum_{\substack{\beta'=1 \\ (\beta'\neq \beta)}}^M\sigma_{u_\beta-u_{\beta'}}(\ry_\beta-\ry_{\beta'})S^{\ry\ry}_{\beta\beta'}.
\end{align}
\end{subequations}
Here $S_{\alpha\alpha'}^{xx}$, $S_{\alpha\beta}^{xy}$, and $S_{\beta\beta'}^{yy}$ are the permutation operators acting on $\{e^{\rx_\alpha}\}$ and $\{e^{\ry_\beta}\}$:
\begin{itemize}
	\item $\rx_\alpha S_{\alpha\alpha'}^{\rx\rx}=S^{\rx\rx}_{\alpha\alpha'}\rx_{\alpha'}$, \quad $\rx_{\alpha'} S_{\alpha\alpha'}^{\rx\rx}=S^{\rx\rx}_{\alpha\alpha'}\rx_{\alpha}$,
	\item $\rx_\alpha S_{\alpha\beta}^{\rx\ry}=S^{\rx\ry}_{\alpha\beta}\ry_\beta$, \quad $\ry_\beta S_{\alpha\beta}^{\rx\ry}=S^{\rx\ry}_{\alpha\beta}\rx_\alpha$,
	\item $\ry_\beta S_{\beta\beta'}^{\ry\ry}=S^{\ry\ry}_{\beta\beta'}\ry_{\beta'}$, \quad $\ry_{\beta'} S_{\beta\beta'}^{\ry\ry}=S^{\ry\ry}_{\beta\beta'}\ry_{\beta}$.
\end{itemize}
Here we show the Dunkl operators defined in \eqref{Dunkl} are pairwise commuting: 
\begin{align}
	\left[d_\alpha^\rx,d_{\alpha'}^\rx\right]
	&=\left[\frac{\partial}{\partial \rx_\alpha},k\sum_{l=1,l\neq \alpha'}^N\sigma_{t_{\alpha'}-t_l}(\rx_{\alpha'}-\rx_l)S^{\rx\rx}_{\alpha'l}+\sum_{\beta=1}^{M}\sigma_{t_{\alpha'}-u_\beta}(\rx_{\alpha'}-\ry_\beta)S^{\rx\ry}_{\alpha'\beta}\right] \nonumber \\
	&\qquad+\left[k\sum_{l=1,l\neq \alpha}^N\sigma_{t_\alpha-t_l}(\rx_\alpha-\rx_l)S^{\rx\rx}_{\alpha l}+\sum_{\beta=1}^{M}\sigma_{t_\alpha-u_\beta}(\rx_\alpha-\ry_\beta)S^{\rx\ry}_{\alpha\beta},\frac{\partial}{\partial \rx_{\alpha'}}\right] \nonumber \\
	&=\left[\frac{\partial}{\partial \rx_\alpha},k\sigma_{t_{\alpha'}-t_\alpha}(\rx_{\alpha'}-\rx_\alpha)S^{\rx\rx}_{\alpha\alpha'}\right]+\left[k\sigma_{t_\alpha-t_{\alpha'}}(\rx_\alpha-\rx_{\alpha'})S^{\rx\rx}_{\alpha\alpha'},\frac{\partial}{\partial \rx_{\alpha'}}\right]\nonumber\\
	&=k\frac{\partial}{\partial \rx_\alpha}\sigma_{t_{\alpha'}-t_\alpha}(\rx_{\alpha'}-\rx_\alpha)S^{\rx\rx}_{\alpha\alpha'}-k\sigma_{t_{\alpha'}-t_\alpha}(\rx_{\alpha'}-\rx_\alpha)\frac{\partial}{\partial \rx_{\alpha'}}\frac{\partial}{\partial \rx_\alpha}S^{\rx\rx}_{\alpha\alpha'}\nonumber\\
	&\qquad+ k\sigma_{t_\alpha-t_{\alpha'}}(\rx_\alpha-\rx_{\alpha'})\frac{\partial}{\partial \rx_\alpha}S^{\rx\rx}_{\alpha\alpha'}-\frac{\partial}{\partial \rx_{\alpha'}}k\sigma_{t_\alpha-t_{\alpha'}}(\rx_\alpha-\rx_{\alpha'})S^{\rx\rx}_{\alpha\alpha'}\nonumber \\
	&=k\left[\frac{\partial}{\partial \rx_\alpha},\sigma_{t_{\alpha'}-t_\alpha}(\rx_{\alpha'}-\rx_\alpha)\right]S^{\rx\rx}_{\alpha\alpha'}+k\left[\frac{\partial}{\partial \rx_{\alpha'}},\sigma_{t_{\alpha'}-t_\alpha}(\rx_{\alpha'}-\rx_\alpha)\right]S^{\rx\rx}_{\alpha\alpha'}=0.
\end{align}
We use \eqref{sigma minus} for the 4th equal sign.
Similarly for the other combinations:
\begin{subequations}
\begin{align}
\left[d_\alpha^\rx,d_\beta^\ry\right]
&=\left[\frac{\partial}{\partial \rx_\alpha},k\sigma_{u_\beta-t_\beta}(\ry_\beta-\rx_\alpha)S_{\alpha\beta}^{\rx\ry}\right]
+\left[\sigma_{t_\alpha-u_\beta}(\rx_\alpha-\ry_\beta)S_{\alpha\beta}^{\rx\ry},k\frac{\partial}{\partial \ry_\beta}\right] \nonumber\\
&=k\left[\frac{\partial}{\partial \rx_\alpha},\sigma_{u_\beta-t_\alpha}(\ry_\beta-\rx_\alpha)\right]S_{\alpha\beta}^{\rx\ry}
+k\left[\sigma_{t_\alpha-u_\beta}(\rx_\alpha-\ry_\beta),\frac{\partial}{\partial \ry_\beta}\right]S_{\alpha\beta}^{\rx\ry}=0, \\
\left[d_\beta^\ry,d_{\beta'}^\ry\right]
&=\left[k\frac{\partial}{\partial \ry_\beta},\sigma_{u_{\beta'}-u_\beta}(\ry_{\beta'}-\ry_{\beta})S_{\beta\beta'}^{\ry\ry}\right]
+\left[\sigma_{u_\beta-u_{\beta'}}(\ry_\beta-\ry_{\beta'})S_{\beta\beta'}^{\ry\ry},k\frac{\partial}{\partial \ry_{\beta'}}\right] \nonumber \\
&=k\left[\frac{\partial}{\partial \ry_\beta},\sigma_{u_{\beta'}-u_\beta}(\ry_{\beta'}-\ry_{\beta})\right]S_{\beta\beta'}^{\ry\ry}
+k\left[\frac{\partial}{\partial \ry_{\beta'}},\sigma_{u_\beta-u_{\beta'}}(\ry_\beta-\ry_{\beta'})\right]S_{\beta\beta'}^{\ry\ry}.
\end{align}
\end{subequations}
For later convenience of calculating conserved commuting Hamiltonians, we will use the combined coordinates $\{\bx_j\}_{j=1}^P$ denoted by:
\begin{equation}
	\bx_j=
	\begin{cases}
		\rx_j & j=1,\dots,N, \\
		\ry_{j-N} & j=N+1,\dots,N+M.
	\end{cases}
\end{equation}
We also define the parity:
\begin{equation}
p(j)=
\begin{cases}
0 & j=1,\dots,N, \\
1 & j=N+1,\dots,N+M. \\
\end{cases}
\end{equation}
Thus one may rewrite Dunkl operators in eq.~\eqref{Dunkl} into a single compact formula for all $j\in[P]=[N+M]$ 
\begin{equation}
d_j=k^{p(j)}\frac{\partial}{\partial \bx_j}+\sum_{l=1,l\neq j}^Pk^{1-p(l)}\sigma_{\bt_j-\bt_l}(\bx_j-\bx_l)S_{jl}; \quad \bx_iS_{ij}=S_{ij}\bx_j.
\end{equation}
The conserved charges are now given as: 
\begin{equation}
\begin{aligned}
&L^{(r)}=\sum_{j=1}^Pk^{-p(j)}(d_j)^r.
\end{aligned}
\end{equation}
Since $d_j$s are commuting, it is easy to see that $L^{(r)}$ are also pairwise commuting
\begin{equation}
\left[L^{(r)},L^{(s)}\right]=0,\quad\forall r,s=1,\dots,P.
\end{equation}
In particular, to recover the original edCM Hamiltonian, we consider:
\begin{equation}
\begin{aligned}
L^{(2)}
=&\sum_{i=1}^Pk^{-p(j)}(d_i)^2=\sum_{\alpha=1}^N(d_\alpha^\rx)^2+\frac{1}{k}\sum_{\beta=1}^M(d_\beta^\ry)^2 \\
=&\sum_{\alpha=1}^N\frac{\partial^2}{\partial \rx_\alpha^2}+k\sum_{\alpha\neq \alpha'}\frac{\partial}{\partial\rx_\alpha}\sigma_{t_\alpha-t_{\alpha'}}(\rx_\alpha-\rx_{\alpha'})+k^2\sum_{\alpha\neq \alpha'}\sigma_{t_\alpha-t_{\alpha'}}(\rx_\alpha-\rx_{\alpha'})\sigma_{t_{\alpha'}-t_{\alpha}}(\rx_{\alpha'}-\rx_{\alpha}) \\
&\qquad\qquad+\sum_{\alpha=1}^N\sum_{\beta=1}^M\frac{\partial}{\partial \rx_\alpha}\sigma_{t_\alpha-u_\beta}(\rx_\alpha-\ry_\beta)+\sigma_{t_\alpha-u_\beta}(\rx_\alpha-\ry_\beta)\sigma_{t_\alpha-u_\beta}(\ry_\beta-\rx_\alpha) \\
+&\frac{k^2}{k}\sum_{\beta=1}^M\frac{\partial^2}{\partial \ry_\beta^2}+\frac{k}{k}\sum_{\beta\neq\beta'}\frac{\partial}{\partial\ry_\beta}\sigma_{u_\beta-u_{\beta'}}(\ry_\beta-\ry_{\beta'})+\frac{1}{k}\sum_{\beta\neq\beta'}\sigma_{u_{\beta}-u_{\beta'}}(\ry_{\beta}-\ry_{\beta'})\sigma_{u_{\beta'}-u_{\beta}}(\ry_{\beta'}-\ry_{\beta}) \\
&\qquad\qquad+\frac{k^2}{k}\sum_{\alpha=1}^N\sum_{\beta=1}^M\frac{\partial}{\partial \rx_\alpha}\sigma_{t_\alpha-u_\beta}(\rx_\alpha-\ry_\beta)+\sigma_{t_\alpha-u_\beta}(\rx_\alpha-\ry_\beta)\sigma_{u_\beta-t_\alpha}(\ry_\beta-\rx_\alpha).
\end{aligned}
\end{equation}
We got Derivatives and product of $\sigma$-function.
In the limit of all $t_\alpha$ and $u_\beta$ are equal, we may use the 4th and 5th properties of $\sigma_t(z)$-function \eqref{sigma properties}  to obtain the edCM Hamiltonian \eqref{DCM}
\begin{equation}
L^{(2)}=2\hat{H} .
\end{equation}
We thus established the quantum integrability of the edCM system.

\subsection{Gauge Origami and Localization}
\paragraph{}
Here we review the relevant details about the so-called gauge origami construction which is an extension of ADHM construction of gauge instantons, 
more details can be found in \cite{Nikita:II,Nikita:III,NaveenNikita}
(See also~\cite{Jeong:2017pai}).
Let start with four complex planes with coordinates $\{z_a\}$, $a=1,2,3,4$ and consider picking two out of them such that the six possible copies  $\bbC^2_A\subset\mathbb{C}^4$ are denoted by the following  double index notation:
\begin{equation}
A\in\{(12),(13),(14),(23),(24),(34)\}=\underline{6}.
\end{equation}
Define the complement of $A$ as $\bar{A}=\{1,2,3,4\}\backslash A$, for instance if $A=(12)$, $\bar{A}=(34)$.
We can thus use $\bbC_{\bA}^2$ to denote the complementary two complex planes transverse to $\bbC_A^2$, such that the total space
$\bbC^4 = \bbC_A^2 \oplus  \bbC_{\bA}^2$.
One can imagine that the six copies of sub-spaces $\bbC_A^2$ are sitting on the six edges of a tetrahedral and its four faces are labeled by the four complex coordinates $\{z_a\}$,
hence the name  ``\begin{CJK}{UTF8}{ipxm}折紙\end{CJK}~(origami)''.
This construction is motivated by the intersecting D-brane configurations using D1-D5-$\overline{\rm D5}$ branes \cite{NaveenNikita}, here we consider the following D(-1)-D3-$\overline{\rm D3}$ intersecting configuration which can be obtained via T-duality transformations:
\begin{center}
\begin{tabular}{|c|c|c|c|c|c|c|c|c|c|c|c|}
	\hline
	Brane Type & \# of Branes & 1 & 2& 3& 4& 5& 6& 7& 8& 9& 10 \\ \hline\hline
	D(-1) & $\k$ & & & & & & & & & & \\ \hline
	$\text{D3}_{(12)}$ & $n_{12}$ & x& x& x& x& & & & & & \\ \hline
	$\overline{\text{D3}}_{(13)}$ & $n_{13}$ & x& x& & & x& x& & & & \\ \hline
	$\text{D3}_{(14)}$ & $n_{14}$ & x& x& & & & & x& x& & \\ \hline
	$\text{D3}_{(23)}$ & $n_{23}$ & & & x& x& x& x& & & & \\ \hline
	$\overline{\text{D3}}_{(24)}$ & $n_{24}$ & & & x& x& & & x& x& & \\ \hline
	$\text{D3}_{(34)}$ & $n_{34}$ & & & & & x& x& x& x& & \\ \hline
\end{tabular}
\end{center}
We labeled each stack of $n_A$ D3 or $\overline{\text{D3}}$ branes by D3$_{A}$ or $\overline{\text{D3}}_{A}$ indicating its four dimension world volume is in $\bbC_A^2$, and gives $U(n_A)$ gauge group.
Notice that the presence of two out of six anti-D3 branes $\overline{\text{D3}}$ is necessary for this intersecting brane configuration to partially preserve $\frac{1}{16}$ of  supersymmetries or two supercharges.
We also introduced $\k$ D(-1) branes, which will play the role of ``spiked instantons'' in this configuration.
Similar to ADHM construction, each gauge group is associated to a vector space $\bN_A=\mathbb{C}^{n_A}$, and one additional vector space $\bK=\mathbb{C}^\k$ is associated to $\k$ D(-1) branes.
The analogous maps acting on $\{\bN_A\}$ and $\bK$ are:
\begin{subequations}
\begin{align}
& I_A: \bN_A\to \bK; \\
& J_A: \bK\to \bN_A; \\
& B_a: \bK\to \bK;\quad a = 1, 2, 3, 4,
\end{align}
\end{subequations}
and we can understand these from the world volume theory of $\k$ D(-1) branes.
Here $\{I_A\}$ and $\{J_A\}$ are the bi-fundamental fields arising from the open string stretching among the D3$_{A}$/$\overline{\text{D3}}_{A}$ and D(-1) branes, while $B_a$ are the complex $U(\k)$ adjoint fields whose diagonal entries label the positions of $\k$ D(-1) branes in the transverse $\bbC^4$.
The analogous real moment map $\mu_{\mathbb{R}}$ and complex moment map $\mu_{\mathbb{C}}$ equations to ADHM construction in $\mathbb{C}^4$ can thus be identified respectively with the so-called D-term, E-term and J-term conditions \cite{NaveenNikita}. Starting with the D-term, which gives the real momentum map $\mu_{\bbR}$
\begin{equation}\label{mu=0}
\{\mu_{\mathbb{R}}=\sum_{a\in\underline{4}}[B_a,B_a^\dagger]+
\sum_{A\in\underline{6}}I_AI_A^\dagger+J_A^\dagger J_A=\zeta\cdot1_\k\}/U(\k), \quad \zeta > 0.
\end{equation}
Here in such an intersecting D-brane configuration, we also turn on constant background NS-NS B-field, it generates the FI parameter in the D(-1) brane world volume theory.
Next we would like to discuss the E- and J-term conditions together. To do so, for each $\bN_A$, let us define the following combinations:
\begin{equation}
\mu_A=[B_a,B_b]+I_AJ_A;\quad a, b \in A
\end{equation} 
and we define $s_A$ as:
\begin{equation}\label{}
s_A=\mu_A+\varepsilon_{A\bar{A}}\mu^\dagger_{\bar{A}},
\end{equation}
where $\varepsilon_{A\bar{A}}$ is a four indices totally antisymmetric tensor ranging over $A$ and its complement $\bA$. 
The analogue to the complex momentum maps are now given by:
\begin{equation}\label{sa=0}
\{s_A=0\}/U(\k).
\end{equation}
Notice that while $\mu_A = 0$ can encode six complex equations, however $s_A$ consist both $\mu_A$ and $\mu^\dagger_{\bar{A}}$ which are mapped into each other under hermitian conjugation, 
there are therefore only six real equations encoded in \eqref{sa=0}.
The reason of using $s_A$ instead of $\mu_A$ is because $s_A$ gives the correct number of degree of freedom, we will show this in a moment. 
In addition, there are equations which do not exist in the usual ADHM construction:
\begin{equation}\label{BIBJ=0}
\{\sigma_{\bar{a}A}=B_{\bar{a}}I_A+\varepsilon_{\bar{a}\bar{b}} B^\dagger_{\bar{b}}J^\dagger_A=0\}/U(\k):\bN_A\to \bK,
\end{equation}
where $\bar{a} \in \bA$ denotes the single index contained in the double index $\bA$.
For every $A$, there are two such equations. 
These equations \eqref{BIBJ=0} appear when one considers the D(-1) and intersecting D3 instanton configurations~\cite{Nikita:II}. 
Now we would like to claim that equations in \eqref{sa=0}, \eqref{mu=0}, and \eqref{BIBJ=0} are sufficient to fix the solution uniquely by showing the number of degrees of freedom and the number of conditions are equal.
Let us start with counting the real degrees of freedom:
\begin{enumerate}
	\item $I_A$: $\sum_A2\times\kappa\times N_A$ real d.o.f.
	\item $J_A$: $\sum_A2\times\kappa\times N_A$ real d.o.f.
	\item $B_a$: $4\times 2\times\kappa^2$ real d.o.f.
\end{enumerate}
which together precisely equals to the number of the conditions:
\begin{enumerate}
	\item Eq.~\eqref{sa=0}: $6\times\kappa^2$ real conditions
	\item Eq.~\eqref{mu=0}: $\kappa^2$ real condition
	\item Eq.~\eqref{BIBJ=0}: $\sum_A2\times2\times\kappa\times N_A$ real conditions
	\item $U(\kappa)$ Symmetry: $\kappa^2$ real condition.
\end{enumerate}
Hence we may show that the dimensions of moduli space defined by $\mathcal{M}_\kappa=\{(\vec{B},\vec{I},\vec{J})\mid\eqref{mu=0},\eqref{sa=0},\eqref{BIBJ=0}\}$  is 
\begin{align}
    \sum_A2\times\kappa\times N_A+\sum_A2\times\kappa\times N_A+8\times\kappa^2-6\times\kappa^2-\kappa^2-\sum_A4\times\kappa\times N_A-\kappa^2=0.
\end{align}
Essentially $\mathcal{M}_\kappa$ only consists of only discrete points. Comparing to ADHM construction which has a moduli space of dimensions $4\kappa N$, additional eq.~\eqref{BIBJ=0} reduces instanton moduli space to be zero dimensional. 
\paragraph{}
However, there also exist open strings stretching between D3-D3 branes which gives additional maps/fields from D-brane construction. These terms are not related to instanton and thus not being considered when constructing instanton moduli space. For instance, when one considers the D(-1)-D3-brane realization of ADHM construction, the open strings with both ends attached to D3 branes are not taken into account. Here we also consider the open strings stretching between $\text{D3}_A$-$\text{D3}_{\bar{A}}$ branes and $\overline{\text{D3}}_A$-$\overline{\text{D3}}_{\bA}$, giving rise to the following conditions:
\begin{equation}\label{JIIJ=0}
\Upsilon_A=J_{\bar{A}}I_A-I^\dagger_{\bar{A}}J^\dagger_A=0:\bN_A\to \bN_{\bar{A}},
\end{equation}
these act as the transversality conditions \cite{Nikita:II}.
The matrices $\{s_A\}, \{\sigma_{\bar{a}A}\}$ and $\{\Upsilon_A\}$ in \eqref{sa=0}, \eqref{BIBJ=0}, and \eqref{JIIJ=0} need to be subjected to the following matrix consistency identity \cite{Nikita:II}:
\begin{equation}\label{Consistency1}
	\sum_{A\in\underline{6}}\text{Tr}(s_As_A^\dagger)+\sum_{A\in\underline{6},\bar{a}\in\underline{4}}\text{Tr}(\sigma_{\bar{a}A}\sigma_{\bar{a}A}^\dagger)+\sum_{A\in\underline{6}}{\rm Tr}(\Upsilon_A\Upsilon_A^\dagger)
	=2\sum_{A\in\underline{6}}\left(\|\mu_A\|^2+\|J_{\bar{A}}I_A \|^2\right)+\sum_{A\in\underline{6},\bar{a}\in\bar{A}}(\|B_{\bar{a}}I_A\|^2+\|J_AB_{\bar{a}}\|^2),
\end{equation}
where $\|\mu_A\|^2=\text{Tr}\left(\mu_A\mu_A^\dagger\right)$. 
By setting each term in the LHS of \eqref{Consistency1} vanishes using  \eqref{sa=0}, \eqref{BIBJ=0}, and \eqref{JIIJ=0}, we can deduce the following constraints:
\begin{subequations}
\begin{align}
	&\{s_A=0\}/U(\k)\implies \mu_A=0,\label{Improv1} \\
	&\{\sigma_{\bar{a}A}=0\}/U(\k)\implies B_{\bar{a}}I_A=0;\quad J_AB_{\bar{a}}=0,\label{Improv2} \\
	&\{\Upsilon_A=0\}/U(\k) \implies J_{\bar{A}}I_A=0,\label{Improv3}
\end{align}
\end{subequations}
which are equivalent to the E- and J-term constraints considered in \cite{NaveenNikita}.
\paragraph{}
It is known that the combination of imposing $\zeta>0$  and dividing by $U(\kappa)$ in \eqref{mu=0} is equivalent to replacing D-term equation \eqref{mu=0} by the  stability condition \cite{Nikita:II}, which states that for any subspace $\bK'\subset\bK$, such that $I_A(\bN_A)$ for all $A\in\underline{6}$ and $B_a\bK'\subset\bK$ for all $a=1,2,3,4$, coincides with $\bK$, i.e. $\bK'=\bK$. 
In other words, 
\begin{equation} \label{stability}
\bK=\sum_A\mathbb{C}[B_1,B_2,B_3,B_4] I_A(\bN_A)/ GL(\bK).
\end{equation}
The equations \eqref{Improv2} and \eqref{Improv3} further shows that $\bK$ can be decomposed into
\begin{equation}\label{Decompose K}
\bK=\bigoplus_A\bK_A;\quad \bK_A=\mathbb{C}[B_a,B_b]I_A(\bN_A),
\end{equation}
The equation \eqref{Decompose K} is essentially the stability condition for familiar ADHM construction. Combining \eqref{Improv1} and \eqref{Decompose K},  we have shown that gauge origami is actually six independent copies of ADHM construction of instantons. 
Finally, the moduli space is now defined as
\begin{equation}\label{moduli}
\mathcal{M}_\kappa(\vec{n})=\{(\vec{B},\vec{I},\vec{J}) \mid \eqref{Improv1},\eqref{stability}\}/\!/ GL(\bK)
\end{equation}
\paragraph{}
There is a symmetry \eqref{sa=0}, \eqref{mu=0}, \eqref{BIBJ=0}, and \eqref{JIIJ=0} enjoys, and thus a symmetry of the moduli space \eqref{moduli}: we can multiply $B_a$ by a phase $B_a\mapsto q_aB_a$, and compensate with $J_A\mapsto q_AJ_A$, $q_A=q_aq_b$ for $A=(ab)$ as long as the product of $q_a$ is subject to:
\begin{align}\label{product of q}
    \prod_{a=1}^4q_a=1.
\end{align}
If we view $\mathbf{q}={\rm diag}(q_1,q_2,q_3,q_4)$ as diagonal matrix, it belongs to the maximal torus $U(1)^3_\epsilon$ of the group $SU(4)$ rotating the $\mathbb{C}^4$.
In the ADHM construction in four dimensions, we usually consider $SO(4)$ rotation acting on $\mathbb{R}^4$, whose maximal torus $U(1)^2$ give rise to two generic $\Omega$-background parameters for complex momentum map. In the gauge origami, if we start with $SO(8)$ rotation acting on $\mathbb{R}^8=\mathbb{C}^4$ with maximal torus $U(1)^4$, one might expect four generic $\Omega$-background parameters. However  conditions defining the moduli space \eqref{sa=0}, \eqref{BIBJ=0}, and \eqref{JIIJ=0} are real equations, which removes over all $U(1)$ phase rotation \eqref{product of q}, leaving maximal torus $U(1)^3$, which preserves some supersymmetry that act
\begin{align}
    \mathbf{q}\cdot[B_a,I_A,J_A]=[q_aB_a,I_A,q_AJ_A].
\end{align}
As stated, the gauge origami can be viewed as a composition of six copies of ADHM instanton constructions. 
Each sub-instanton vector space $\bK_A$ has its fixed-points labeled by a set of Young diagrams $\vec{\lambda}_{A}=(\lambda^{(1)}_A,\dots,\lambda^{(n_A)}_A)$, each Young diagram is labeled by $\lambda^{(\alpha)}_A=(\lambda_{A,\alpha,1},\lambda_{A,\alpha,2},\dots)$, $\alpha=1,\dots,n_A$, such that:
\begin{equation}
\bK_A=\bigoplus_{\alpha=1}^{n_A}\bK_{A,\alpha};\quad \bK_{A,\alpha}=\bigoplus_{i=1}^{\ell({\lambda_{A,\alpha})}}\bigoplus_{j=1}^{\lambda_{A,\alpha,i}}B_a^{i-1}B_b^{j-1}(I_A e_{A,\alpha}); \quad \bN_A=\mathbb{C}^{n_A}=\bigoplus_{\alpha=1}^{n_A}\mathbb{C}e_{A,\alpha}.
\end{equation} where $e_{A,\alpha}$ is the fixed basis of the vector space $\mathbb{C}^{n_A}$.
We will also use $\underline{\lambda}=\{\vec{\lambda}_A\}$ to denote the set of all gauge origami Young diagrams.
\paragraph{}
As in the usual ADHM construction, we denote the character on each $\bN_A$ and $\bK_A$ as:
\begin{equation}
\begin{aligned}
N_A :=\sum_{\alpha=1}^{n_A}e^{a_{A,\alpha}};\quad
K_A :=\sum_{\alpha=1}^{n_A}e^{a_{A,\alpha}}\sum_{(i,j)\in\lambda_{A}^{(\alpha)}}q_a^{i-1}q_b^{j-1}.
\end{aligned}
\end{equation}
The character on the tangent space of the moduli space defined in eq.~\eqref{moduli} can be written as
\begin{equation}
\cT_{\underline{\lambda}}=\mathcal{N}\mathcal{K}^*-P_1P_2P_3\mathcal{K}\mathcal{K}^*-q_AN_A^*N_{\bar{A}},
\end{equation}
with the following definition
\begin{equation}
\mathcal{N}=\sum_{A\in\underline{6}}P_{\bar{A}}N_A;\qquad \mathcal{K}=\sum_{A\in\underline{6}}K_A,
\end{equation}
and the following notation:
\begin{equation}
q_a=e^{\epsilon_a};\quad P_a=1-q_a;\quad q_A=q_aq_b;\quad P_A=P_aP_b.
\end{equation}
Using \eqref{product of q}, it shows $(\epsilon_a)_{a=1,\ldots,4}$ are subject to the constraint:
\begin{equation}
\sum_{a=1}^4\epsilon_a=0.
\label{eq:epsilons_constraint}
\end{equation}

\textbf{Example 1: }
Consider all $n_A\equiv0$ except $n_{12}=N$, the character is given as
\begin{equation}\label{n12=N}
\begin{aligned}
\mathcal{T}_{\lambda_{12}}
& =\mathcal{N}\mathcal{K}^*-P_1P_2P_3\mathcal{K}\mathcal{K}^* \\
& =(1-q_3)(1-q_4)N_{12}K^*_{12}-P_1P_2(1-q_3)K_{12}K^*_{12} \\
& =(1-q_3-q_4-q_3q_4)N_{12}K^*_{12}-P_1P_2(1-q_3)K_{12}K^*_{12} \\
& =(1-q_3)[N_{12}K^*_{12}+qN_{12}^*K_{12}-P_1P_2K_{12}K^*_{12}],
\end{aligned}
\end{equation}
Define the operation $\mathbb{E}$ as:
\begin{equation}
	\mathbb{E}\left[\sum_{i\in I_+}e^{W^+_{i}}-\sum_{i\in I_-}e^{W^-_{i}}\right]=\frac{\prod_{i\in I_-}(W^-_{i})}{\prod_{i\in I_+}(W^+_{i})}.
\end{equation}
We found that the instanton partition function of 4d $U(N)$ $\mathcal{N}=2^*$ theory defined in \eqref{N=2* IP} can be obtained from this character:
\begin{align}
    Z_\text{inst}[\vec{\lambda}=\vec{\lambda}_{12}] = \mathbb{E} \left[ \mathcal{T}_{\lambda_{12}}\right]
\end{align}
under the identification of the adjoint mass $m=\epsilon_3$.
\paragraph{}
\textbf{Example 2: }
Consider all $n_A\equiv0$ except $n_{12}=N$, $n_{34}=1$. 
Take $N_{34}=e^x$, and we have
\begin{equation}
\begin{aligned}
\mathcal{T}_{\lambda_{12},\lambda_{34}}
=&\mathcal{N}\mathcal{K}^*-P_1P_2P_3\mathcal{K}\mathcal{K}^* \\
=&[(1-q_3)(1-q_4)N_{12}+(1-q_1)(1-q_2)N_{34}](K_{12}+K_{34})^*-P(1-q_3)(K_{12}+K_{34})(K_{12}+K_{34})^* \\
=&(1-q_3)[N_{12}K^*_{12}+qN_{12}^*K_{12}-PK_{12}K^*_{12}]+(1-q_1)[N_{34}K^*_{34}+q_3q_4N_{34}^*K_{34}-P_3P_4K_{34}K^*_{34}] \\
&+(1-q_3)(1-q_4)N_{12}K_{34}^*+(1-q_1)(1-q_2)N_{34}K_{12}^*-P(1-q_3)(K_{12}K_{34}^*+K_{34}K_{12}^*).
\end{aligned}
\end{equation}
Comparing with $\mathbb{X}(x)$ in eq.~\eqref{Xbb-obs}, we realize that
\begin{equation}
\mathbb{X}(x)[\mu=\lambda_{34}]=\mathbb{E}\left[(1-q_3)(1-q_4)N_{12}K_{34}^*+(1-q_1)(1-q_2)N_{34}K_{12}^*-P(1-q_3)(K_{12}K_{34}^*+K_{34}K_{12}^*)\right]
\end{equation}
is the fundamental $qq$-character of $\widehat{A}_0$ quiver, with $m=\epsilon_3$ and $-m-\epsilon=\epsilon_4$
a.k.a. the crossed instanton configuration \cite{Nikita:II}. For $n_{34}>1$, one obtains higher order  qq-character of $\hat{A}_0$ quiver \cite{Nikita:I}.

\section{Elliptic Double Calogero-Moser from Gauge Origami}\label{eDCM-Origami}
\paragraph{}
Now we would like to see how a special case of the gauge origami construction reviewed earlier is naturally connected with the edCM system.
Let us consider a special case with only two stacks of overlapping D3 branes, i.e.
\begin{equation}
    n_{12}=N,\qquad n_{23}=M,
\end{equation}
while all the remaining $n_{A \neq (12), (23)}=0$.
The Young diagrams associated with such a gauge origami configuration are denoted as:
\be
	\vec{\lambda}_{12}=(\lambda_{12}^{(1)},\dots,\lambda_{12}^{(N)});\quad\vec{\lambda}_{23}=(\lambda_{23}^{(1)},\dots,\lambda_{23}^{(M)}),
\ee
with each individual Young diagram represented by
\begin{equation}
	\lambda_{12}^{(\alpha)}=(\lambda_{12,\alpha,1},\lambda_{12,\alpha,2},\dots);\quad \lambda_{23}^{(\beta)}=(\lambda_{23,\beta,1},\lambda_{23,\beta,2},\dots)
\end{equation}
where $\alpha=1,\dots,N$, $\beta=1,\dots,M$. For each gauge group, we can define the following combinations:
\begin{subequations}
\begin{align}
	&x_{\alpha i}=a_\alpha+(i-1)\epsilon_1+\lambda_{12,\alpha,i}\epsilon_2;\quad x^{(0)}_{\alpha i}=a_\alpha+(i-1)\epsilon_1,  \\
	&x_{\beta j}=b_{\beta}+(j-1)\epsilon_3+\lambda_{23,\beta,j}\epsilon_2
	;\quad x^{(0)}_{\beta j}=b_{\beta}+(j-1)\epsilon_3.
\end{align}
\end{subequations}
{This special configuration is also called the folded instanton \cite{Nikita:II,Nikita:V,Koroteev:2019byp}.} The partition function of such a gauge origami configuration is given by: 
\begin{equation}\label{Znm-origami}
\begin{aligned}
	\mathcal{Z}_{\text{inst}}
	& =\sum_{\{\vec{\lambda}_{12}\}}\sum_{\{\vec{\lambda}_{23}\}}\mathfrak{q}^{|\vec{\lambda}_{12}|+|\vec{\lambda}_{23}|}\mathcal{Z}_{11}[\vec{\lambda}_{12}]\mathcal{Z}_{33}[\vec{\lambda}_{23}]\mathcal{Z}_{13}[\vec{\lambda}_{12},\vec{\lambda}_{23}]\mathcal{Z}_{31}[\vec{\lambda}_{23},\vec{\lambda}_{12}],
\end{aligned}
\end{equation}
where
\begin{subequations}
\begin{align}
\mathcal{Z}_{11}[\vec{\lambda}_{12}]
=\prod_{(\alpha i)\neq(\alpha' i')}& \frac{\Gamma(\epsilon_2^{-1}(x_{\alpha i}-x_{\alpha' i'}-\epsilon_1))}{\Gamma(\epsilon_2^{-1}(x_{\alpha i}-x_{\alpha' i'}))}\cdot\frac{\Gamma(\epsilon_2^{-1}(x_{\alpha i}-x_{\alpha' i'}-\epsilon_3))}{\Gamma(\epsilon_2^{-1}(x_{\alpha i}-x_{\alpha' i'}-\epsilon_1-\epsilon_3))} \nonumber\\
&  \times\frac{\Gamma(\epsilon_2^{-1}(x_{\alpha i}^{(0)}-x_{\alpha' i'}^{(0)}))}{\Gamma(\epsilon_2^{-1}(x_{\alpha i}^{(0)}-x_{\alpha' i'}^{(0)}-\epsilon_1))}\cdot\frac{\Gamma(\epsilon_2^{-1}(x_{\alpha i}^{(0)}-x_{\alpha' i'}^{(0)}-\epsilon_1-\epsilon_3))}{\Gamma(\epsilon_2^{-1}(x_{\alpha i}^{(0)}-x_{\alpha' i'}^{(0)}-\epsilon_3))}, \\
\mathcal{Z}_{33}[\vec{\lambda}_{23}]
=\prod_{(\beta j)\neq(\beta' j')}& \frac{\Gamma(\epsilon_2^{-1}(x_{\beta j}-x_{\beta' j'}-\epsilon_1))}{\Gamma(\epsilon_2^{-1}(x_{\beta j}-x_{\beta' j'}))}\cdot\frac{\Gamma(\epsilon_2^{-1}(x_{\beta j}-x_{\beta' j'}-\epsilon_3))}{\Gamma(\epsilon_2^{-1}(x_{\beta j}-x_{\beta' j'}-\epsilon_1-\epsilon_3))} \nonumber\\
&  \times\frac{\Gamma(\epsilon_2^{-1}(x_{\beta j}^{(0)}-x_{\beta' j'}^{(0)}))}{\Gamma(\epsilon_2^{-1}(x_{\beta j}^{(0)}-x_{\beta' j'}^{(0)}-\epsilon_1))}\cdot\frac{\Gamma(\epsilon_2^{-1}(x_{\beta j}^{(0)}-x_{\beta' j'}^{(0)}-\epsilon_1-\epsilon_3))}{\Gamma(\epsilon_2^{-1}(x_{\beta j}^{(0)}-x_{\beta' j'}^{(0)}-\epsilon_3))}, \\
\mathcal{Z}_{13}[\vec{\lambda}_{12},\vec{\lambda}_{23}]
=\prod_{(\alpha i)}\prod_{(\beta j)}& \frac{\Gamma(\epsilon_2^{-1}(x_{\alpha i}-x_{\beta j}-\epsilon_1))}{\Gamma(\epsilon_2^{-1}(x_{\alpha i}-x_{\beta j}))}\cdot\frac{\Gamma(\epsilon_2^{-1}(x_{\alpha i}-x_{\beta j}-\epsilon_3))}{\Gamma(\epsilon_2^{-1}(x_{\alpha i}-x_{\beta j}-\epsilon_1-\epsilon_3))} \nonumber\\
&  \times\frac{\Gamma(\epsilon_2^{-1}(x_{\alpha i}^{(0)}-x_{\beta j}^{(0)}))}{\Gamma(\epsilon_2^{-1}(x_{\alpha i}^{(0)}-x_{\beta j}^{(0)}-\epsilon_1))}\cdot\frac{\Gamma(\epsilon_2^{-1}(x_{\alpha i}^{(0)}-x_{\beta j}^{(0)}-\epsilon_1-\epsilon_3))}{\Gamma(\epsilon_2^{-1}(x_{\alpha i}^{(0)}-x_{\beta j}^{(0)}-\epsilon_3))}, \\
\mathcal{Z}_{31}[\vec{\lambda}_{23},\vec{\lambda}_{12}]
=\prod_{(\beta j)}\prod_{(\alpha i)}& \frac{\Gamma(\epsilon_2^{-1}(x_{\beta j}-x_{\alpha i}-\epsilon_1))}{\Gamma(\epsilon_2^{-1}(x_{\beta j}-x_{\alpha i}))}\cdot\frac{\Gamma(\epsilon_2^{-1}(x_{\beta j}-x_{\alpha i}-\epsilon_3))}{\Gamma(\epsilon_2^{-1}(x_{\beta j}-x_{\alpha i}-\epsilon_1-\epsilon_3))} \nonumber\\
&  \times\frac{\Gamma(\epsilon_2^{-1}(x_{\beta j}^{(0)}-x_{\alpha i}^{(0)}))}{\Gamma(\epsilon_2^{-1}(x_{\beta j}^{(0)}-x_{\alpha i}^{(0)}-\epsilon_1))}\cdot\frac{\Gamma(\epsilon_2^{-1}(x_{\beta j}^{(0)}-x_{\alpha i}^{(0)}-\epsilon_1-\epsilon_3))}{\Gamma(\epsilon_2^{-1}(x_{\beta j}^{(0)}-x_{\alpha i}^{(0)}-\epsilon_3))}.
\end{align}
\end{subequations}
We take NS limit $\epsilon_2\to0$ while keeping $\epsilon_1$ and $\epsilon_3$ fixed. 
Following the similar procedures in Section~\ref{sec:BAEfromInstanton}, one finds the saddle point configuration satisfies 
\begin{equation}\label{Q-dCM}
	1+\mathfrak{q}\frac{Q(x_{\gamma k}-\epsilon_4)Q(x_{\gamma k}-\epsilon_3)Q(x_{\gamma k}-\epsilon_1)}{Q(x_{\gamma k}+\epsilon_4)Q(x_{\gamma k}+\epsilon_3)Q(x_{\gamma k}+\epsilon_1)}=0;\quad (\gamma k)=\{(\alpha i),(\beta j)\},
\end{equation}
where
\begin{equation}\label{Qp}
Q(x)=\prod_{\alpha=1}^N\prod_{i=1}^\infty(x-x_{\alpha i})\prod_{\beta=1}^M\prod_{j=1}^\infty(x-x_{\beta j}).
\end{equation}
We denote the Young diagrams which satisfy \eqref{Q-dCM} the limit shape configurations $\vec{\lambda}_{12}^*$ and $\vec{\lambda}_{23}^*$, they dominate the full folded instanton partition function given in \eqref{Znm-origami} in NS limit:
\begin{align}
    \mathcal{Z}_{\text{inst}}
	& \approx\mathfrak{q}^{|\vec{\lambda}_{12}^*|+|\vec{\lambda}_{23}^*|}\mathcal{Z}_{11}[\vec{\lambda}_{12}^*]\mathcal{Z}_{33}[\vec{\lambda}_{23}^*]\mathcal{Z}_{13}[\vec{\lambda}_{12}^*,\vec{\lambda}_{23}^*]\mathcal{Z}_{31}[\vec{\lambda}_{23}^*,\vec{\lambda}_{12}^*]=\mathfrak{q}^{|\vec{\lambda}_{12}^*|+|\vec{\lambda}_{23}^*|}\mathcal{Z}_\text{inst}[\vec{\lambda}_{12}^*,\vec{\lambda}_{23}^*].
\end{align}
To find the resultant BAE, we consider the twisted superpotential arising in the NS limit: $\cW=\lim_{\epsilon_2\to0}[\epsilon_2\cZ]=\cW_\text{classical}+\cW_\text{1-loop}+\cW_\text{inst}$, whose equation of motion is now given by:
\begin{equation}
\frac{1}{2\pi i}\frac{\partial \mathcal{W}(g_\gamma)}{\partial g_{\gamma}}=n_{\gamma};\qquad n_\gamma\in\mathbb{Z},
\end{equation} 
with $g_\gamma\in\{a_{\alpha},b_{\beta}\}$, $g_\gamma=a_\alpha$ for $\gamma=1, \dots, N$ and $g_\gamma=b_\beta$ for $\gamma=N+1, \dots N+M$. The classical twisted superpotential is given as
\begin{align}
    \cW_\text{classical}=-\log\mathfrak{q}
    \sum_{\alpha=1}^N\frac{a_\alpha^2}{2\epsilon_1}-\log\mathfrak{q}\sum_{\beta=1}^{M}\frac{b_\beta^2}{2\epsilon_3},
\end{align}
and the perturbative one-loop twisted superpotential is
\begin{align}
    \cW_\text{1-loop}=\frac{1}{2}\sum_{(\gamma k)\neq(\gamma' k')}
    &\{ f(x_{\gamma i
    k}^{(0)}-x_{\gamma' k'}^{(0)}-\epsilon_1)-f(x_{\gamma k}^{(0)}-x_{\gamma' k'}^{(0)}+\epsilon_1) \nonumber\\
    & +f(x_{\gamma k}^{(0)}-x_{\gamma' k'}^{(0)}-\epsilon_3)-f(x_{\gamma k}^{(0)}-x_{\gamma' k'}^{(0)}+\epsilon_3) \nonumber\\
    &+f(x_{\gamma k}^{(0)}-x_{\gamma' k'}^{(0)}+\epsilon_4)-f(x_{\gamma k}-x_{\gamma' k'}^{(0)}-\epsilon_4)\},
\end{align}
with $x_{\gamma k}^{(0)}\in\{x_{\alpha i}^{(0)},x_{\beta j}^{(0)}\}$.
The BAE now can be obtained after some elaborated calculations, following the same procedures as in \eqref{How to Gamma}: 
\begin{subequations}\label{BAE-dCM}
\begin{align}
	1&=\mathfrak{q}^{-\frac{a_\alpha}{2\epsilon_1}}
	\prod_{\alpha'(\neq\alpha)}\frac{\Gamma\left(\frac{a_\alpha-a_{\alpha'}}{\epsilon_1}\right)}{\Gamma\left(-\frac{a_\alpha-a_{\alpha'}}{\epsilon_1}\right)}\frac{\Gamma\left(\frac{-\epsilon_3-(a_\alpha-a_{\alpha'})}{\epsilon_1}\right)}{\Gamma\left(\frac{-\epsilon_3+a_\alpha-a_{\alpha'}}{\epsilon_1}\right)}
	\prod_{\beta}\frac{\Gamma\left(\frac{a_\alpha-b_\beta}{\epsilon_3}\right)}{\Gamma\left(-\frac{a_\alpha-b_\beta}{\epsilon_3}\right)}\frac{\Gamma\left(\frac{-\epsilon_1-(a_\alpha-b_\beta)}{\epsilon_3}\right)}{\Gamma\left(\frac{-\epsilon_1+a_\alpha-b_\beta}{\epsilon_3}\right)}, \\
	1&=\mathfrak{q}^{-\frac{b_\beta}{2\epsilon_3}}
	\prod_{\alpha}\frac{\Gamma\left(\frac{b_\beta-a_\alpha}{\epsilon_1}\right)}{\Gamma\left(-\frac{b_\beta-a_\alpha}{\epsilon_1}\right)}\frac{\Gamma\left(\frac{-\epsilon_3-(b_\beta-a_\alpha)}{\epsilon_1}\right)}{\Gamma\left(\frac{-\epsilon_3+b_\beta-a_\alpha}{\epsilon_1}\right)}
	\prod_{\beta'(\neq\beta)}\frac{\Gamma\left(\frac{b_\beta-b_{\beta'}}{\epsilon_3}\right)}{\Gamma\left(-\frac{b_\beta-b_{\beta'}}{\epsilon_3}\right)}\frac{\Gamma\left(\frac{-\epsilon_1-(b_\beta-b_{\beta'})}{\epsilon_3}\right)}{\Gamma\left(\frac{-\epsilon_1+b_\beta-b_{\beta'}}{\epsilon_3}\right)}.
\end{align}
\end{subequations}
Comparing with the BAE of eCM in eq.~\eqref{BAEA0}, eq.~\eqref{BAE-dCM} consists of two copies of the eCM systems.
To the best of our knowledge, the BAE for the edCM system has not appeared in the literature, 
we therefore propose that \eqref{BAE-dCM} is a possible one.
We will provide supporting evidence this statement by deriving the commuting Hamiltonians explicitly from the folded instanton configuration in the following section.

\subsection{$\bbX(x)$ for Elliptic Double Calogero-Moser System}
\paragraph{}
As we have shown in the previous sections that, $\mathbb{X}$-function was constructed upon auxiliary lattice as an enhanced version of the original $T$-function, and it is the characteristic polynomial for the eCM system. We would like to see if similar construction also applies for the edCM system, using the gauge origami partition function.
\paragraph{}
We claim that the resultant $\mathbb{X}(x)$ should be of the following factorizable form:
\begin{equation}\label{Xbb-obs2}
\mathbb{X}(x)=\mathbb{X}_1(x)\times\mathbb{X}_3(x) .
\end{equation}
When we restore $\epsilon_2$ dependence, the two factors are:
\begin{equation}
\begin{aligned}
& \mathbb{X}_1(x)=\sum_{\{\mu_1\}}\frac{Q(x+\epsilon_1)}{Q(x)}\mathfrak{q}^{|\mu_1|}B_{12}[\mu_{1}]\prod_{(\bi,\bj)\in\mu_1}\frac{Q(x+s_{1,\bi\bj}-\epsilon_4)Q(x+s_{1,\bi\bj}-\epsilon_3)Q(x+s_{1,\bi\bj}-\epsilon_1)}{Q(x+s_{1,\bi\bj}+\epsilon_4)Q(x+s_{1,\bi\bj}+\epsilon_3)Q(x+s_{1,\bi\bj}+\epsilon_1)}, \\
& \mathbb{X}_3(x)=\sum_{\{\mu_3\}}\frac{Q(x+\epsilon_3)}{Q(x)}\mathfrak{q}^{|\mu_3|}B_{32}[\mu_3]\prod_{(\bi,\bj)\in\Lambda^\mu_3}\frac{Q(x+s_{3,\bi\bj}-\epsilon_4)Q(x+s_{3,\bi\bj}-\epsilon_3)Q(x+s_{3,\bi\bj}-\epsilon_1)}{Q(x+s_{3,\bi\bj}+\epsilon_4)Q(x+s_{3,\bi\bj}+\epsilon_3)Q(x+s_{3,\bi\bj}+\epsilon_1)}, \\
\end{aligned}
\end{equation}
with the parameters given by:
\begin{subequations}
\begin{align}
	& s_{1,\bi\bj}=\bi\epsilon_3+\bj\epsilon_4; \quad {B}_{12}[{\mu_1}]=\prod_{(\bi,\bj)\in\mu_1}\left[1+\frac{\epsilon_1\epsilon_2}{(\epsilon_3(l_{\bi\bj}+1)-\epsilon_4 a_{\bi\bj})(\epsilon_3(l_{\bi\bj}+1)-\epsilon_4 a_{\bi\bj}+\epsilon_1+\epsilon_2)}\right],\label{B12} \\
	& s_{3,\bi\bj}=\bi\epsilon_1+\bj\epsilon_4 ;\quad {B}_{32}[{\mu_3}]=\prod_{(\bi,\bj)\in\mu_3}\left[1+\frac{\epsilon_3\epsilon_2}{(\epsilon_1(l_{\bi\bj}+1)-\epsilon_4 a_{\bi\bj})(\epsilon_1(l_{\bi\bj}+1)-\epsilon_4 a_{\bi\bj}+\epsilon_3+\epsilon_2)}\right], \label{B32}
\end{align}
\end{subequations}
and the constraint \eqref{eq:epsilons_constraint} applies.
Comparing with the gauge origami construction, we see that $\mathbb{X}_1(x)$ corresponds to the configuration $n_{12}=N$, $n_{23}=M$, and $n_{34}=1$, with $\mu_1=\lambda_{34}$, while $\mathbb{X}_3(x)$ corresponds to the configuration $n_{12}=N$, $n_{23}=M$, and $n_{14}=1$, with $\mu_3=\lambda_{14}$. 
The $x$-independent terms in $\mathbb{X}_1(x)$ (or $\mathbb{X}_3(x)$) can be viewed as auxiliary instanton partition of $U(1)$ gauge theory living on $\mathbb{C}_3\times\mathbb{C}_4$ (or $\mathbb{C}_1\times\mathbb{C}_4$) four-dimensional subspace. 
Here we would like to stress that $\mathbb{X}(x)$ is not equivalent to having a single gauge origami consisting $n_{12}=N$, $n_{23}=M$, $n_{34}=1$, $n_{14}=1$, rather a product of two different gauge origami systems.
The configuration with $n_{12}=N$, $n_{23}=M$, $n_{34}=1$, $n_{14}=1$ can only be factorizable under NS limit, and without orbifolding. If either orbifolding is implemented or we keep $\epsilon_2$ finite, it is not factorizable.
\paragraph{}
To show that $\mathbb{X}(x)$ has the correct degree $P$, let us define the analogous functions to eq.~\eqref{Y-func}:
\begin{equation}
Y_1(x)=\frac{Q_1(x)}{Q_1(x-\epsilon_1)},\quad Q_1(x)=\prod_{\alpha=1}^N\prod_{i=1}^\infty(x-x_{\alpha i});\quad Y_3(x)=\frac{Q_3(x)}{Q_3(x-\epsilon_3)},\quad Q_3(x)=\prod_{\beta=1}^M\prod_{j=1}^\infty(x-x_{\beta j}).
\end{equation}
We may now take large $x$ limit for both $Y_1(x)$ and $Y_3(x)$ and find:
\begin{subequations}
\begin{align}
& Y_1(x)\approx \prod_{\alpha=1}^N\prod_{i=1}^\infty\frac{(x-x_{\alpha i}^{(0)})}{(x-x_{\alpha i}^{(0)}-\epsilon_1)}=\prod_{\alpha=1}^N(x-a_\alpha)\approx x^N, \\
& Y_3(x)\approx \prod_{\beta=1}^M\prod_{j=1}^\infty\frac{(x-x_{\beta j}^{(0)})}{(x-x_{\beta j}^{(0)}-\epsilon_3)}=\prod_{\beta=1}^M(x-b_\beta)\approx x^M.
\end{align}
\end{subequations}
Following the similar argument given for the ordinary eCM model, one can prove that $\mathbb{X}(x)$ is analytic in the complex plane.
We may now rewrite $\mathbb{X}(x)$ using $Y_1(x)$ and $Y_3(x)$:
\begin{equation}
\begin{aligned}
\mathbb{X}(x)
=&\sum_{\{\mu_1,\mu_3\}}\left[{Y_1(x+\epsilon_1)}\prod_{n=1}^\infty\frac{Y_3(x+n\epsilon_3)}{Y_3(x+\epsilon_1+n\epsilon_3)}\mathfrak{q}^{|\mu_1|}B[\mu_1]\prod_{(\bi,\bj)\in\mu_1}\frac{Q(x+s_{1,\bi\bj}-\epsilon_4)Q(x+s_{1,\bi\bj}-\epsilon_3)Q(x+s_{1,\bi\bj}-\epsilon_1)}{Q(x+s_{1,\bi\bj}+\epsilon_4)Q(x+s_{1,\bi\bj}+\epsilon_3)Q(x+s_{1,\bi\bj}+\epsilon_1)}\right] \\
&\times\left[{Y_3(x+\epsilon_3)}\prod_{n=1}^\infty\frac{Y_1(x+n\epsilon_1)}{Y_1(x+\epsilon_3+n\epsilon_1)}\mathfrak{q}^{|\mu_3|}B[\mu_3]\prod_{(\bi,\bj)\in\mu_3}\frac{Q(x+s_{3,\bi\bj}-\epsilon_4)Q(x+s_{3,\bi\bj}-\epsilon_3)Q(x+s_{3,\bi\bj}-\epsilon_1)}{Q(x+s_{3,\bi\bj}+\epsilon_4)Q(x+s_{3,\bi\bj}+\epsilon_3)Q(x+s_{3,\bi\bj}+\epsilon_1)}\right].
\end{aligned}
\end{equation}
In large $x$ limit, we have
\begin{equation}
Y_1(x+\epsilon_1)Y_3(x+\epsilon_3)\approx x^Nx^M=x^{P},
\end{equation}
which is the desired degree.
We will next show that $\bbX(x)$ indeed reproduces the commuting Hamiltonians of the edCM system.

\subsection{Commuting Hamiltonians from $\mathbb{X}(x)$ for edCM system}
\paragraph{}
Here we again introduce $\mathbb{Z}_P$-type full surface defect on $\mathbb{C}_{13}^2\subset\mathbb{C}^4$ with orbifolding in $\mathbb{C}_{24}^2\subset\mathbb{C}^4$, such that the orbifolding acts on the coordinate of $\mathbb{C}^4$ by $(\mathbf{z}_1,\mathbf{z}_2,\mathbf{z}_3,\mathbf{z}_4)\to(\mathbf{z}_1,\zeta \mathbf{z}_2,\mathbf{z}_3,\zeta^{-1}\mathbf{z}_4)$ with $\zeta^P=1$. This is not a co-dimension two defect, but rather it should be interpreted as a generalization of the surface defects. Similar to what we had done for Toda and eCM systems, we define the coloring function on the indices of moduli parameters $c:\{\alpha\}_{\alpha=1}^N\cup\{\beta\}_{\beta=1}^M\to\mathbb{Z}_P$ which assigns each color $\alpha$ and $\beta$ to a representation $R_\omega$ of $\mathbb{Z}_P$, $\omega=0,1,\dots,P-1$.
In the simplest case, $c$ is defined as
\begin{equation}\label{dCM coloring}
	c(\alpha)=\alpha-1;\quad c(\beta)=N+\beta-1. 
\end{equation}
For this coloring function, we will denote
$[\alpha]=\{0,\dots,N-1\}$ and $[\beta]=\{N,\dots,N+M-1\}$ such that $[\alpha]\cup[\beta]=\{0,\dots,P-1\}$, which is the range of the index $\omega$. 
Orbifolding also splits coupling $\mathfrak{q}$ into $P$-copies denoted by
\begin{equation}
	\mathfrak{q}=\prod_{\omega=0}^{P-1}\mathfrak{q}_\omega;\quad \mathfrak{q}_\omega=\frac{\mathfrak{z}_\omega}{\mathfrak{z}_{\omega-1}},
\end{equation}
with
\begin{equation}
	\mathfrak{z}_\omega=
	\begin{cases}
		z_\alpha=e^{\rx_\alpha}, & c^{-1}(\omega+1)=\alpha\in\{\alpha=1,\dots,N\} \\
		w_\beta=e^{\ry_\beta}, & c^{-1}(\omega+1)=\beta\in\{\beta=1,\dots,M\}.
	\end{cases}
\end{equation}
Under the orbifolding, we have
\begin{equation}
	Y_1(x)=\prod_{\omega=1}^{P-1}Y_{1,\omega}(x);\quad Y_3(x)=\prod_{\omega=0}^{P-1}Y_{3,\omega}(x),
\end{equation}
with
\begin{subequations}
\begin{align}
	Y_{1,\omega}(x)
	&=(x-a_{\omega})\prod_{(\alpha,(i,j))\in K^{12}_{\omega}}
	\left[\frac{(x-a_\alpha-(i-1)\epsilon_1-\epsilon_1)}{(x-a_\alpha-(i-1)\epsilon_1)}\right]\prod_{(\alpha,(i,j))\in K^{12}_{\omega+1}}
	\left[\frac{(x-a_\alpha-(i-1)\epsilon_1)}{(x-a_\alpha-(i-1)\epsilon_1-\epsilon_1)}\right], \\
	Y_{3,\omega}(x)
	&=(x-b_{\omega})\prod_{(\beta,(i,j))\in K^{32}_{\omega}}
	\left[\frac{(x-b_\beta-(i-1)\epsilon_3-\epsilon_3)}{(x-b_\beta-(i-1)\epsilon_1}\right]\prod_{(\beta,(i,j))\in K^{32}_{\omega+1}}
	\left[\frac{(x-b_\beta-(i-1)\epsilon_3)}{(x-b_\beta-(i-1)\epsilon_3-\epsilon_3)}\right],
\end{align}
\end{subequations}
under NS limit $\epsilon_2\to0$ while keeping $\epsilon_1$ and $\epsilon_3$ finite.
We also consider the following Young diagram boxes under orbifolding:
\begin{subequations}\label{K-double}
\begin{align}
	&K^{12}_\omega:=\{(\alpha,(i,j)) \mid \alpha=1,\dots,N;\quad(i,j)\in\lambda_{12}^{(\alpha)};\quad c(\alpha)+j\equiv\omega \ \text{mod} \ P\}, \\
	&K^{32}_\omega:=\{(\beta,(i,j)) \mid \beta=1,\dots,M ;\quad(i,j)\in\lambda^{(\beta)}_{32};\quad c(\beta)+j\equiv\omega \ \text{mod} \ P\},
\end{align}
\end{subequations}
where $K_\omega^{12}$ and $K_\omega^{32}$ are the collections of Young diagram boxes from $\vec{\lambda}_{12}$ and $\vec{\lambda}_{32}$ which are assigned to the representation $R_\omega$ under orbifolding. They are the same as the definition in eq.~\eqref{K}. 
Denoting
\begin{subequations}\label{kv-data-double}
\begin{align}
	&k^{12}_\omega=|K^{12}_\omega|,\quad\nu^{12}_\omega=k^{12}_{\omega}-k^{12}_{\omega+1},\quad \sigma_\omega^{12}=\frac{\epsilon_1}{2}k_\omega^{12}+\sum_{(\alpha,(i,j))\in K^{12}_\omega}(a_\alpha+(i-1)\epsilon_1), \\
	&k^{32}_\omega=|K^{32}_\omega|,\quad\nu^{32}_\omega=k^{32}_{\omega}-k^{32}_{\omega+1};\quad \sigma_\omega^{32}=\frac{\epsilon_3}{2}k_\omega^{32}+\sum_{(\beta,(i,j))\in K^{32}_\omega}(b_\beta+(i-1)\epsilon_3),
\end{align}
\end{subequations}
as the generalization to eq.~\eqref{K} and eq.~\eqref{kv-data}. Performing the large $x$ expansion of $Y_{1, \omega}(x)$ and $Y_{3, \omega}(x)$ under orbifolding gives 
\begin{subequations}
\begin{align}
	& Y_{1,\omega}(x)=[{x-a_{c^{-1}(\omega)}}]\exp\left[\frac{\epsilon_1}{x}\nu^{12}_{\omega-1}+\frac{\epsilon_1}{x^2}(\sigma^{12}_{\omega-1}-\sigma^{12}_\omega)+\cdots\right];\qquad c(\omega)\in[\alpha], \\
	& Y_{3,\omega}(x)=[{x-b_{c^{-1}(\omega)}}]\exp\left[\frac{\epsilon_3}{x}\nu^{32}_{\omega-1}+\frac{\epsilon_3}{x^2}(\sigma^{32}_{\omega-1}-\sigma^{32}_\omega)+\cdots\right];\qquad c(\omega)\in[\beta].
\end{align}
\end{subequations}
The notation here follows eq.~\eqref{kv-data-double}. 
Under orbifolding, $\mathbb{X}_1(x)$ and $\mathbb{X}_3(x)$ now splits into
\begin{subequations}
\begin{align}
\mathbb{X}_{1,\omega}(x)
=&Y_{1,\omega+1}(x+\epsilon_1)\prod_{n=1}^\infty\frac{Y_{3,\omega}(x+n\epsilon_3)}{Y_{3,\omega}(x+\epsilon_1+n\epsilon_3)}\sum_{\{\mu_1\}}\mathbb{B}^{12}_\omega[\mu_1]\times \nonumber\\
&\prod_{(\bi,\bj)\in\mu_1}\frac{Y_{1,\omega+1-\bj}(x+s_{\bi\bj}-\epsilon_3)Y_{1,\omega+1-\bj+1}(x+s_{\bi\bj}-\epsilon_4)}{Y_{1,\omega+1-\bj}(x+s_{\bi\bj})Y_{1,\omega+1-\bj+1}(x+s_{\bi\bj}+\epsilon_1)}\frac{Y_{3,\omega+1-\bj}(x+s_{\bi\bj}-\epsilon_1)Y_{3,\omega+1-\bj+1}(x+s_{\bi\bj}-\epsilon_4)}{Y_{3,\omega+1-\bj}(x+s_{\bi\bj})Y_{3,\omega+1-\bj+1}(x+s_{\bi\bj}+\epsilon_3)}, \\
\mathbb{X}_{3,\omega}(x)
=&Y_{3,\omega+1}(x+\epsilon_3)\prod_{n=1}^\infty\frac{Y_{1,\omega}(x+n\epsilon_1)}{Y_{1,\omega}(x+\epsilon_3+n\epsilon_1)}\sum_{\{\mu_3\}}\mathbb{B}^{32}_\omega[\mu_3]\times \nonumber\\
&\prod_{(\bi,\bj)\in\mu_3}\frac{Y_{1,\omega+1-\bj}(x+s_{\bi\bj}-\epsilon_3)Y_{1,\omega+1-\bj+1}(x+s_{\bi\bj}-\epsilon_4)}{Y_{1,\omega+1-\bj}(x+s_{\bi\bj})Y_{1,\omega+1-\bj+1}(x+s_{\bi\bj}+\epsilon_1)}\frac{Y_{3,\omega+1-\bj}(x+s_{\bi\bj}-\epsilon_1)Y_{3,\omega+1-\bj+1}(x+s_{\bi\bj}-\epsilon_4)}{Y_{3,\omega+1-\bj}(x+s_{\bi\bj})Y_{3,\omega+1-\bj+1}(x+s_{\bi\bj}+\epsilon_3)}.
\end{align}
\end{subequations}
Here $\mathbb{B}_\omega^{12}[\mu_1]$ and $\mathbb{B}_\omega^{32}[\mu_3]$ are the $U(1)$ orbifolded instanton partitions living on $\mathbb{C}_3\times\mathbb{C}_4$ and $\mathbb{C}_1\times\mathbb{C}_4$ with instanton configuration $\mu_1$ and $\mu_3$ respectively, where
\begin{subequations}
\begin{align}
	& \mathbb{B}^{12}_\omega[\mu_1]=\left.\prod_{(\bi,\bj)\in\mu_1}\mathfrak{q}_{\omega+1-\bj}B_{1}(\epsilon_3 l_{\bi\bj})\right|_{a_{\bi\bj}=0}=\prod_{l=1}^{\mu_{1,1}}\prod_{h=1}^{\mu^T_{1,l}-\mu^T_{1,l+1}}\frac{\mathfrak{z}_\omega}{\mathfrak{z}_{\omega-l}}B_{1}(\epsilon_3h); \quad B_1(x)=1+\frac{\epsilon_1}{x},\label{B^12}\\
	& \mathbb{B}^{32}_\omega[\mu_3]=\left.\prod_{(\bi,\bj)\in\mu_3}\mathfrak{q}_{\omega+1-\bj}B_{3}(\epsilon_1 l_{\bi\bj})\right|_{a_{\bi\bj}=0}=\prod_{l=1}^{\mu_{3,1}}\prod_{h=1}^{\mu^T_{3,l}-\mu^T_{3,l+1}}\frac{\mathfrak{z}_\omega}{\mathfrak{z}_{\omega-l}}B_{3}(\epsilon_1h); \quad B_3(x)=1+\frac{\epsilon_3}{x}.\label{B^32}
\end{align}
\end{subequations}
We will consider the summation over all possible partition
\begin{subequations}\begin{align}
\mathbb{B}_\omega^{12} & = \sum_{\{\mu_1\}} \mathbb{B}_\omega^{12}[\mu_1], \\
\mathbb{B}_\omega^{32} & = \sum_{\{\mu_3\}} \mathbb{B}_\omega^{32}[\mu_3],
\end{align}
\end{subequations}
we can regard $B_1(x)$ and $B_3(x)$ are orbifolded version of \eqref{B12} and \eqref{B32}.
\paragraph{}
After some tedious but similar calculations, one gets for $(\omega+1)\in[\alpha]$, the large $x$ expansion gives
\begin{equation}
\begin{aligned}
	&\frac{1}{\mathbb{B}^{12}_{\omega}}\mathbb{X}_{1,\omega}(x) 
	=x+\epsilon_1-a_{c^{-1}(\omega+1)}+\epsilon_1\nu^{12}_\omega+\frac{1}{x}\Bigg[
    \frac{1}{2}(\epsilon_1\nu^{12}_\omega-a_{c^{-1}(\omega+1)})^2-\frac{1}{2}(a_{c^{-1}(\omega+1)})^2+\epsilon_1D_\omega^{12}+
    \\
	&\left.\sum_{\{\mu_1\}}\frac{\mathbb{B}^{12}_\omega[\mu_1]}{\mathbb{B}^{12}_\omega}\left(\epsilon_3\sum_{(\omega'+1)\in[\alpha]}\epsilon_4k_{1,\omega'}^\mu-\left(\epsilon_1\nu^{12}_{\omega'}-a_{c^{-1}(\omega'+1)}\right)\nu^\mu_{1,\omega'}
	+\epsilon_1\sum_{(\omega'+1)\in[\beta]}\epsilon_4k_{1,\omega'}^\mu-\left(\epsilon_3\nu^{12}_{\omega'}-b_{c^{-1}(\omega'+1)}\right)\nu^\mu_{1,\omega'}\right)\right]+\cdots \\
\end{aligned}
\end{equation}
with $D_\omega^{12}=\sigma_{\omega}^{12}-\sigma_{\omega+1}^{12}$. We divide the $\mathbb{X}_{1,\omega}(x)$-function by the factor $\mathbb{B}_\omega^{12}$ for the normalization. Similarly for $(\omega+1)\in[\beta]$, we have large $x$ expansion:
\begin{equation}
\begin{aligned}
	&\frac{1}{\mathbb{B}^{32}_{\omega}}\mathbb{X}_{3,\omega}(x) =x+\epsilon_3-b_{c^{-1}(\omega+1)}+\epsilon_3\nu^{32}_\omega+\frac{1}{x}\Bigg[
    \frac{1}{2}(\epsilon_3\nu^{32}_\omega-b_{c^{-1}(\omega+1)})^2-\frac{1}{2}(b_{c^{-1}(\omega+1)})^2+\epsilon_3D_\omega^{32}+
    \\
	&\left.\sum_{\{\mu_3\}}\frac{\mathbb{B}^{32}_\omega[\mu_3]}{\mathbb{B}^{32}_\omega}\left(\epsilon_3\sum_{(\omega'+1)\in[\alpha]}\epsilon_4k_{3,\omega'}^\mu-\left(\epsilon_1\nu_{\omega'}^{32}-a_{c^{-1}(\omega'+1)}\right)\nu^\mu_{3,\omega'}
	+\epsilon_1\sum_{(\omega'+1)\in[\beta]}\epsilon_4k_{3,\omega'}^\mu-\left(\epsilon_3\nu_{\omega'}^{32}-b_{c^{-1}(\omega'+1)}\right)\nu^\mu_{3,\omega'}\right)\right]+\cdots \\
\end{aligned}
\end{equation}
with $D_\omega^{32}=\sigma_{\omega}^{32}-\sigma_{\omega+1}^{32}$. Again we normalize $\mathbb{X}_{3,\omega}(x)$ by diving the overall expression with $\mathbb{B}_\omega^{32}$.
For all $\omega=0,\dots,P-1$, we can define:
\begin{equation}
\nabla^\mathfrak{q}=\mathfrak{q}\frac{\partial}{\partial\mathfrak{q}};\quad\nabla^\mathfrak{q}_\omega=\mathfrak{q}_\omega\frac{\partial}{\partial \mathfrak{q}_\omega};\quad\nabla^\mathfrak{z}_\omega=\mathfrak{z}_\omega\frac{\partial}{\partial \mathfrak{z}_\omega}.
\end{equation}
The following combination gives a degree $P$ function of $x$.
\begin{equation}\label{X polynomial double}
	\prod_{(\omega+1)\in[\alpha]}\frac{\mathbb{X}_{1,\omega}}{\epsilon_1\mathbb{B}^{12}_\omega}\prod_{(\omega+1)\in[\beta]}\frac{\mathbb{X}_{3,\omega}}{\epsilon_3\mathbb{B}^{32}_\omega},
\end{equation}
this can be seen from the fact that each $\frac{\mathbb{X}_{1, \omega}}{\mathbb{B}_{\omega}^{12}}$ or $\frac{\mathbb{X}_{3, \omega}}{\mathbb{B}_{\omega}^{32}}$ factor in the product above is of degree one.
Denote the first commuting Hamiltonian as
\begin{equation}
h_1=\sum_{(\omega+1)\in[\alpha]}\epsilon_1\nu^{12}_\omega-a_{c^{-1}(\omega+1)}+\sum_{(\omega+1)\in[\beta]}\epsilon_3\nu^{32}_\omega-b_{c^{-1}(\omega+1)}=\sum_{(\omega+1)\in[\alpha]}P^{12}_\omega+\sum_{(\omega+1)\in[\beta]}P^{32}_\omega,
\end{equation}
the conjugated momentum is denoted as $P_\omega^{12}=(\epsilon_1\nabla^{z}_\omega-a_{c^{-1}(\omega+1)})$ when $(\omega+1)\in[\alpha]$, $P_\omega^{23}=(\epsilon_3\nabla^w_\omega-b_{c^{-1}(\omega+1)})$ when $(\omega+1)\in[\beta]$.  
\paragraph{}
We may also write the second commuting Hamiltonian $h_2$ as
\begin{equation}
\begin{aligned}
	h_2
	=&\sum_{(\omega+1)\in[\alpha]}\frac{1}{2\epsilon_1}(P_\omega^{12})^2-\frac{1}{2\epsilon_1}(a_{c^{-1}(\omega+1)})^2 +\sum_{(\omega+1)\in[\beta]}\frac{1}{2\epsilon_3}(P_\omega^{23})^2-\frac{1}{2\epsilon_3}(b_{c^{-1}(\omega+1)})^2 \\
	&+k\sum_{(\omega+1)\in[\alpha]}(\epsilon_4\nabla^\mathfrak{q}_\omega-P_\omega^{12}\nabla^\mathfrak{z}_\omega)\log\mathbb{B}^{12}_{\alpha\alpha}(\vec{z};\tau)
	+\sum_{\omega}(\epsilon_4\nabla^\mathfrak{q}_\omega-P_\omega\nabla^\mathfrak{z}_\omega)\log\mathbb{B}^{12}_{\alpha\beta}(\vec{z},\vec{w};\tau) \\
	&+\sum_{\omega}(\epsilon_4\nabla^\mathfrak{q}_\omega-P_\omega\nabla^\mathfrak{z}_\omega)\log\mathbb{B}^{32}_{\beta\alpha}(\vec{w},\vec{z};\tau)
	+\frac{1}{k}\sum_{(\omega+1)\in[\beta]}(\epsilon_4\nabla^\mathfrak{q}_\omega-P_\omega^{23}\nabla^\mathfrak{z}_\omega)\log\mathbb{B}^{32}_{\beta\beta}(\vec{w};\tau), \\
\end{aligned}
\end{equation}
with $k=\epsilon_3/\epsilon_1$.
Let us define:
\begin{equation}
\mathbb{B}^{12}=\prod_{(\omega+1)\in[\alpha]}\mathbb{B}^{12}_\omega=\mathbb{B}^{12}_{\alpha\alpha}(\vec{z};\tau)\mathbb{B}^{12}_{\alpha\beta}(\vec{z},\vec{w};\tau);\quad\mathbb{B}^{32}=\prod_{(\omega+1)\in[\beta]}\mathbb{B}^{32}_\omega=\mathbb{B}^{32}_{\beta\alpha}(\vec{w},\vec{z};\tau)\mathbb{B}^{32}_{\beta\beta}(\vec{w};\tau),
\end{equation}
where the $z_\alpha$ and $w_\beta$ dependent functions are
\begin{align}
	&\mathbb{B}^{12}_{\alpha\alpha'}(\vec{z};\tau)=\left[\prod_{N\geq\alpha>\alpha'\geq0}\frac{1}{1-\frac{z_\alpha}{z_{\alpha'}}}\prod_{\alpha=0}^{N}\prod_{\alpha'=0}^{N}\frac{1}{(\mathfrak{q}\frac{z_\alpha}{z_{\alpha'}};\mathfrak{q})_\infty} \right]^{-\frac{\epsilon_4}{\epsilon_3}}
	;\quad\mathbb{B}^{12}_{\alpha\beta}(\vec{z},\vec{w};\tau)=\left[\prod_{\alpha=1}^N\prod_{\beta=1}^{M}\frac{1}{(\mathfrak{q}\frac{z_\omega}{w_\beta};\mathfrak{q})_\infty} \right]; \nonumber \\
	&\mathbb{B}^{32}_{\beta\alpha}(\vec{w},\vec{z};\tau)=\left[\prod_{\alpha=1}^N\prod_{\beta=1}^M\frac{1}{1-\frac{w_\beta}{z_{\alpha}}}\prod_{\alpha=1}^{N}\prod_{\beta=1}^M\frac{1}{(\mathfrak{q}\frac{w_\beta}{z_\alpha};\mathfrak{q})_\infty} \right]
	;\quad\mathbb{B}^{32}_{\beta\beta'}(\vec{w};\tau)=\left[\prod_{M\geq\beta>\beta'\geq0}\frac{1}{1-\frac{w_\beta}{w_{\beta'}}}\prod_{\beta=1}^M\prod_{\beta'=1}^{M}\frac{1}{(\mathfrak{q}\frac{w_\beta}{w_{\beta'}};\mathfrak{q})_\infty} \right]^{-\frac{\epsilon_4}{\epsilon_1}}.
\end{align}
We remark $\epsilon_4 = - (\epsilon_1 + \epsilon_3)$ in the NS limit $\epsilon_2 \to 0$ due to the constraint \eqref{eq:epsilons_constraint}.
Notice that $\mathbb{B}^{12}_{\alpha\beta}$ is not symmetrical to $\mathbb{B}^{32}_{\beta\alpha}$ for $\mathfrak{q}$-independent part. The reason of this is due to the specific coloring function $c$ we chose in eq.~\eqref{dCM coloring}, and we will now explain how this works. 
Based on \eqref{B^12} and \eqref{B^32} before summing over all Young diagrams, the $\mathfrak{q}$ independent part comes from the product:
\begin{align}
	\prod_{\bj=1}^{\omega'}\mathfrak{q}_{\omega+1-\bj}=\frac{\mathfrak{z}_\omega}{\mathfrak{z}_{\omega-\omega'}};\quad \omega>\omega'\geq1.
\end{align}
Using the coloring function defined in eq.~\eqref{dCM coloring} for $z_\alpha=\mathfrak{z}_{\alpha-1}$ and $w_\beta=\mathfrak{z}_{N+\beta-1}$, there is no such way to have
\begin{align}
    \frac{z_\alpha}{w_\beta}=\frac{\mathfrak{z}_{\alpha-1}}{\mathfrak{z}_{N+\beta-1}}=\frac{1}{\mathfrak{q}_\alpha\mathfrak{q}_{\alpha+1}\cdots\mathfrak{q}_{N+\beta-1}}
\end{align}
since this expression contributes negative number of instantons (inverse power on counting $\mathfrak{q}_\omega$), while its inverse is legit.
This is the cause of asymmetry between $\mathbb{B}^{12}_{\alpha\beta}$ and $\mathbb{B}^{32}_{\beta\alpha}$ in the $\mathfrak{q}$-independent factor.
\paragraph{}
As before we dropped the $z$-independent factors (which appeared in eq.~\eqref{B=QF}), since they can be removed in the final stage by redefining zero point of energy level as shown previously. The $q$-Pochhammer notation is defined in eq.~\eqref{q-Pochhammer}.
We may therefore denote the $z$-dependent parts as:
\begin{equation}
\begin{aligned}
	\log\mathbb{B}^{12}_{\alpha\alpha}=-\frac{\epsilon_4}{\epsilon_3}\log\mathbb{Q}_{\alpha\alpha};\quad \log\mathbb{B}^{12}_{\alpha\beta}\mathbb{B}^{32}_{\beta\alpha}=\log\mathbb{Q}_{\alpha\beta};\quad\log\mathbb{B}^{32}_{\beta\beta}=-\frac{\epsilon_4}{\epsilon_1}\log{\mathbb{Q}_{\beta\beta}},
\end{aligned}
\end{equation}
such that $\mathbb{Q}_{12}$ and $\mathbb{Q}_{32}$ combine to give a full $\theta$-function, i.e. 
\be
\mathbb{Q}_{12}\mathbb{Q}_{32}=\mathbb{Q}_{\alpha\alpha}(\vec{z})\mathbb{Q}_{\alpha\beta}(\vec{z},\vec{w})\mathbb{Q}_{\beta\beta}(\vec{w})\eta(\tau)^P\frac{\mathfrak{q}^{P^2/24}}{\vec{\mathfrak{z}}^{\vec{\rho}}}.
\ee 
with $\vec{\rho}$ now is the $P$-dimensional Weyl vector.
One may refer to how the additional factors appears in eq.~\eqref{Q}.
This structure also shows up in trigonometric limit~\cite{HiJack},with
\begin{subequations}
\begin{align}
    \mathbb{Q}_{\alpha\alpha}^{-1}
    &=\left[\prod_{N\geq\alpha>\alpha'\geq1}\frac{\theta_{11}\left(\frac{z_\alpha}{z_{\alpha'}};\tau\right)}{\eta(\tau)}\right]; \\
    \mathbb{Q}_{\alpha\beta}^{-1}
    &=\left[\prod_{\alpha=1}^N\prod_{\beta=1}^M\frac{\theta_{11}\left(\frac{z_\alpha}{w_{\beta}};\tau\right)}{\eta(\tau)}\right]; \\
    \mathbb{Q}_{\beta\beta}^{-1}
    &=\left[\prod_{M\geq\beta>\beta'\geq1}\frac{\theta_{11}\left(\frac{w_\beta}{w_{\beta'}};\tau\right)}{\eta(\tau)}\right].
\end{align}
\end{subequations}
Following a similar calculation as in the previous section, the potential after canonical transformation \eqref{canonical trans} can be written as
\begin{align}
	{V}=
	&\sum_{(\omega+1)\in[\alpha]}\frac{k\epsilon_4}{2}\left(\nabla_\omega^\mathfrak{z}\right)^2\log\mathbb{Q}_{\alpha\alpha}+\sum_{(\omega+1)\in[\beta]}\frac{\epsilon_4}{2k}\left(\nabla_\omega^\mathfrak{z}\right)^2\log\mathbb{Q}_{\beta\beta} 
	+\frac{\epsilon_4}{2}\sum_{(\omega+1)\in[\alpha]\cup[\beta]}\left(\nabla^\mathfrak{z}_\omega\right)^2\log\mathbb{Q}_{\alpha\beta}
	\nonumber \\
	=&-\epsilon_1k(k+1)\sum_{\alpha>\alpha'}\wp(z_{\alpha}/z_{\alpha'};\tau)-\epsilon_1(k+1)\sum_{\alpha,\beta}\wp(z_\alpha/w_\beta;\tau)-\frac{\epsilon_1}{k}(k+1)\sum_{\beta>\beta'}\wp(w_\beta/w_{\beta'};\tau)
\end{align}
with $\epsilon_4 = -(\eps_1+\eps_3)$. We now have the second Hamiltonian written as:
\begin{align}
-\frac{1}{\epsilon_1}h_2=&\sum_{\alpha=1}^N\frac{1}{2}\left(\nabla_\alpha^z-\frac{a_\alpha}{\epsilon_1}\right)^2+\sum_{\beta=1}^M\frac{k}{2}\left(\nabla_\alpha^w-\frac{b_\beta}{\epsilon_3}\right)^2  \nonumber \\
&-k(k+1)\sum_{\alpha>\alpha'}\wp(z_{\alpha}/z_{\alpha'};\tau)-(k+1)\sum_{\alpha,\beta}\wp(z_\alpha/w_\beta;\tau)-\left(\frac{1}{k}+1\right)\sum_{\beta>\beta'}\wp(w_\beta/w_{\beta'};\tau).
\end{align}
Similar to eCM system, using the fact that \eqref{Xbb-obs2} is the q-character defined upon limit shape, which dominates in the NS-limit.
\begin{align}
    t(x)=\langle\mathcal{X}(x)\rangle=\frac{\mathbb{X}(x)\mathcal{Z}_\text{inst}[\vec{\lambda}_{12}^*,\vec{\lambda}_{23}^*]}{\mathcal{Z}_\text{inst}[\vec{\lambda}_{12}^*,\vec{\lambda}_{23}^*]}.
\end{align}
where $t(x)=x^P+E_1x^{P-1}+E_2x^{P-2}+\cdots+E_P$ is the characteristic polynomial. When Hamiltonian are treated as operators, we have
\begin{align}
    \mathbb{X}(x)\mathcal{Z}_\text{inst}[\vec{\lambda}_{12}^*,\vec{\lambda}_{23}^*](\vec{\rx},\vec{\ry})=t(x)\mathcal{Z}_\text{inst}[\vec{\lambda}_{12}^*,\vec{\lambda}_{23}^*](\vec{\rx},\vec{\ry}),
\end{align}
and the Canonical transformation gives a prefactor to the ground state wave function $\Psi(\bx,\by)$ of the edCM model:
\begin{align}\label{Eigen function of edCM}
    \Psi(\vec{\rx},\vec{\ry})=\left[\mathbb{Q}_{\alpha\alpha}(\vec{\rx})^\frac{\epsilon_1+\epsilon_3}{\epsilon_1}\mathbb{Q}_{\alpha\beta}(\vec{\rx},\vec{\ry})\mathbb{Q}_{\beta\beta}(\vec{\ry})^\frac{\epsilon_1
    +\epsilon_3}{\epsilon_3}\right]^{-1}\mathcal{Z}_\text{inst}[\vec{\lambda}_{12}^*,\vec{\lambda}_{23}^*](\vec{\rx},\vec{\ry});\quad h_2\Psi(\vec{\rx},\vec{\ry})=E_2\Psi(\vec{\rx},\vec{\ry}).
\end{align}
{We have thus successfully reproduced the potential of the edCM system defined in eq.~\eqref{DCM}.
The parameter dictionary can be summarized into the following table:

\begin{center}
	\begin{tabular}{|c|c|c|}
		\hline
		& Gauge Theory & Integrable System \\ \hline\hline
		$a_\alpha$, $b_\beta$ & Coulomb Moduli & Momenta \\ \hline
		$\tau$ & Complex gauge coupling & Elliptic modulus \\ \hline
		$\epsilon_1$, $\epsilon_3$ & $\Omega$-deformation parameters & Coupling constant (in the form of $k={\epsilon_3}/{\epsilon_1}$) \\ \hline
		$N$, $M$ & Gauge group rank & Number of particles \\ \hline
		$z_\alpha$, $w_\beta$ & Ratios between orbifolded couplings & Exponentiated coordinates  \\
		\hline
	\end{tabular}
\end{center}

\section{Discussions}\label{sec:discussion}
\paragraph{}
Let us end this work by discussing a few possible future directions.
\begin{enumerate}
    \item The edCM was shown to have no natural classical limits in Section~\ref{GaugeOrigami}. This also implies that the usual story of identifying the gauge theoretic Seiberg-Witten curve  with the spectral curve of classical integrable system does not apply here. Due to the same reason, the quantum Dunkl operators, rather than the classical Lax matrices, were used to construct commuting Hamiltonians. However we have also shown that we can use intersecting D-brane configuration to construct the gauge-origami theory which are directly related to the edCM systems. It would be very interesting to consider the possible M-theory lift of such a configuration, this should illuminate the construction of the inherently quantum Seiberg-Witten curve of gauge-origami theory hence the spectral curves of the edCM systems. It would be also interesting to explore a direct gauge theoretic interpretation of the (double) Dunkl operator.
    \item Double Calogero-Moser system was first constructed by considering root system of supergroup. When coupling constant $k<0$, the gauge theory associated to the edCM system with Hamiltonian in eq.~\eqref{DCM} should be a supergroup gauge theory, whose partition function is obtained in~\cite{Kimura:2019msw}. We hope to report on this and other related topics in our forthcoming work.
    
    \item The gauge groups we discussed in this paper are of SU-type. In principle one may also consider SO/Sp gauge groups. It will be nice if one can find commuting Hamiltonians of corresponding integrable system using the orbifolding and large $x$ expansion. The same argument also extend to various types of quiver gauge theory (several A-types quiver gauge theory has been considered in \cite{Jeong:2017pai}).
    In addition, we would like to know if the gauge origami construction can be generalized to SO/Sp gauge groups. This will involve introducing orientifolds to the intersecting D-brane construction for SU case.
    \item In the single eCM system,  the quantization condition $m = \epsilon_3 = \mathbb{Z} \times \hbar = \mathbb{Z} \times \epsilon_1$ can be implemented. How does this arrangement affect both the integrable system and gauge theory? And we would like to know whether the edCM shares the same quantization condition?
\end{enumerate}

       


\subsection*{Acknowledgements}
This work of HYC was supported in part by Ministry of Science and Technology
(MOST) through the grant 107 -2112-M-002-008-.
The work of TK was supported in part by JSPS Grant-in-Aid for Scientific Research (No.~JP17K18090), the MEXT-Supported Program for the Strategic Research Foundation at Private Universities ``Topological Science'' (No. S1511006), JSPS Grant-in-Aid for Scientific Research on Innovative Areas ``Topological Materials Science'' (No. JP15H05855), ``Discrete Geometric Analysis for Materials Design'' (No.~JP17H06462), and also by the French ``Investissements d’Avenir'' program, project ISITE-BFC (No.~ANR-15-IDEX-0003). The work of NL is supported by Simons Center for Geometry and Physics and State of New York.
HYC and NL are also grateful to the hospitality of Keio University during the completion of this work. We also would like to thank Saebyeok Jeong, Peter Koroteev, Nikita Nekrasov for commenting our draft when it was being finalized.

\appendix
\section{Appendix }

\subsection{Random Partition}
A partition is defined as a way of expressing a non-negative integer $n$ as summation over other non-negative integers. Each partition can be labeled by a Young diagram $\lambda=(\lambda_1,\lambda_2,\dots,\lambda_{\ell(\lambda)})$ with $\lambda_i\in\mathbb{N}$ such that 
\begin{equation}
n=|\lambda|=\sum_{i=1}^{\ell(\lambda)}\lambda_i.
\end{equation} 
We define the generating function of such a partition as
\begin{subequations}\label{phi}
\begin{align}
&\sum_{\lambda}\mathfrak{q}^{|\lambda|}=\frac{1}{(\mathfrak{q};\mathfrak{q})_\infty},\quad(\mathfrak{q};\mathfrak{q})_\infty=\prod_{n=1}^\infty\left(1-\mathfrak{q}^n\right); \\
&\sum_{\lambda}t^{\ell(\lambda)}\mathfrak{q}^{|\lambda|}=\frac{1}{(\mathfrak{q}t;\mathfrak{q})_\infty};\quad (\mathfrak{q}t;\mathfrak{q})_\infty=\prod_{n=1}^\infty(1-t\mathfrak{q}^n).
\end{align}
\end{subequations}
The $\mathfrak{q}$-shifted factorial (the $\mathfrak{q}$-Pochhammer symbol) is defined as
\begin{equation}\label{q-Pochhammer}
(z;q)_n=\prod_{m=0}^{n-1}(1-zq^m)
\end{equation}

\subsection{Elliptic Function}
\paragraph{}
Here we fix our notation for the elliptic functions. The so-called Dedekind eta function is denoted as
\begin{equation}\label{eta}
\eta(\tau)=e^{\frac{\pi i\tau}{12}}(\mathfrak{q};\mathfrak{q})_\infty .
\end{equation}
The first Jacobi $\theta$ function is denoted as:
\begin{equation}\label{theta}
\theta_{11}(z;\tau)=ie^{\frac{\pi i\tau}{4}}z^{\frac{1}{2}}(\mathfrak{q};\mathfrak{q})_\infty (\mathfrak{q}z;\mathfrak{q})_\infty(z^{-1};\mathfrak{q})_\infty,
\end{equation} 
whose series expansion 
\begin{equation}\label{theta2}
\theta_{11}(z;\tau)=i\sum_{r\in\mathbb{Z}+\frac{1}{2}}(-1)^{r-\frac{1}{2}}z^re^{\pi i\tau r^2}=i\sum_{r\in\mathbb{Z}+\frac{1}{2}}(-1)^{r-\frac{1}{2}}e^{rx}e^{\pi i\tau r^2},
\end{equation}
implies that it obeys the heat equation
\begin{equation}
\frac{1}{\pi i}\frac{\partial}{\partial\tau}\theta_{11}(z;\tau)=(z\partial_z)^2\theta_{11}(z;\tau).
\end{equation}
The Weierstrass $\wp$-function
\be\label{Def:p-function}
\wp(z) = \frac{1}{z^2}+\sum_{p, q \ge 0} \left\{\frac{1}{(z+p+q\tau)^2}-\frac{1}{(p+q\tau)^2}\right\},
\ee 
is related to theta and eta functions by
\begin{equation}
\wp(z;\tau)=-(z\partial_z)^2\log\theta_{11}(z;\tau)+\frac{1}{\pi i}\partial_\tau\log\eta(\tau).
\end{equation}

\subsection{Higher rank Theta function}
Let us define
\begin{equation}
\Theta_{A_{N-1}}(\vec{z};\tau)=\eta(\tau)^{N}\prod_{\alpha>\beta}\frac{\theta_{11}(z_\alpha/z_\beta;\tau)}{\eta(\tau)}
\end{equation}
as the rank $N-1$ theta function, which also satisfies the heat equation \cite{Kac}
\begin{equation}\label{heat}
N\frac{\partial}{\partial\tau}\Theta_{A_{N-1}}(\vec{z};\tau)=\pi i\Delta_{\vec{z}}\Theta_{A_{N-1}}(\vec{z};\tau),
\end{equation}
with the $N$-variable Laplacian:
\begin{align}
    \Delta_{\vec{z}}=\sum_{\omega=0}^{N-1}(z_\omega\partial_{z_\omega})^2.
\end{align}

\subsection{Orbifolded Partition}
\paragraph{}
For the purpose in the main text, we consider the orbifolded coupling 
\begin{align}
    \mathfrak{q}=\prod_{\omega=0}^N\mathfrak{q}_\omega;\quad \mathfrak{q}_{\omega+N}=\mathfrak{q}_\omega,
\end{align}
and
\begin{align}
    \mathfrak{q}_\omega=\frac{z_\omega}{z_{\omega-1}};\quad z_{\omega+N}=\mathfrak{q}z_{\omega}.
\end{align}
We also consider the orbifolded version of the generating function of partitions $(\mathfrak{q};\mathfrak{q})_\infty^{-1}$ in \eqref{phi}. Given a finite partition $\lambda=(\lambda_1,\dots,\lambda_{\ell(\lambda)})$, we define
\begin{equation}
\mathbb{Q}^{\lambda}_\omega
=\prod_{j=1}^{\lambda_1}\mathfrak{q}_{\omega+1-j}^{\lambda_j^t}
=\prod_{i=1}^{\ell(\lambda)}\frac{z_\omega}{z_{\omega-\lambda_{i}}},
\end{equation}
where we used the relation \eqref{q-z}.
The summation over all possible partition is given by
\begin{equation}\label{Q-form}
\mathbb{Q}_\omega=\sum_{\lambda}\mathbb{Q}^{\lambda}_\omega=\sum_{\lambda}\prod_{i=1}^{\ell(\lambda)}\left(\frac{z_\omega}{z_{\omega-\lambda_{i}}}\right)=\sum_{l_0,\dots,l_{N-1},l\geq0}\prod_{\alpha=1}^{N-1}\left(\frac{z_\omega}{z_{\alpha}}\right)^{l_\alpha}\mathfrak{q}^l.
\end{equation}
The function $\mathbb{Q}(\vec{z};\tau)$ is the orbifolded version of the generating function of partitions \eqref{phi}, 
\begin{align}\label{Q}
\mathbb{Q}
&=\prod_{\omega=0}^{N-1}\mathbb{Q}_\omega(\vec{z};\tau) \nonumber\\
&=\prod_{N-1 \geq \alpha>\beta \geq 0}\frac{1}{(\frac{z_\alpha}{z_\beta};\mathfrak{q})_\infty(\mathfrak{q}\frac{z_\beta}{z_\alpha};\mathfrak{q})_\infty}\prod_{\alpha=0}^{N-1}\frac{1}{(\mathfrak{q};\mathfrak{q})_\infty} \nonumber\\
&=\prod_{N-1\geq\alpha>\beta\geq0}\frac{\mathfrak{q}^{1/12}\eta(\tau)\sqrt{z_\alpha/z_\beta}}{\theta_{11}(z_\alpha/z_\beta;\tau)}\times\left[\frac{\mathfrak{q}^{1/24}}{\eta(\tau)}\right]^N \nonumber\\
&=\left[\eta(\tau)^{-N}\prod_{N-1\geq\alpha>\beta\geq0}\frac{\eta(\tau)}{\theta_{11}(z_\alpha/z_\beta;\tau)}\right]\frac{\mathfrak{q}^{N^2/24}}{\vec{z}^{\vec{\rho}}} \nonumber\\
&=\frac{1}{\Theta_{A_{N-1}}(\vec{z};\tau)}\frac{\mathfrak{q}^{N^2/24}}{\vec{z}^{\vec{\rho}}},
\end{align}
where $\vec{\rho}$ is the Weyl vector of $SU(N)$ Lie group, whose entries are given as
\begin{equation}
\vec{\rho}=(\rho_0,\dots,\rho_{N-1});\quad\rho_\omega=\omega-\frac{N-1}{2};\quad |\vec{\rho}|^2=\sum_{\omega=0}^{N-1}\rho_\omega^2=\frac{N(N^2-1)}{12};\quad\vec{z}^{\vec{\rho}}=\prod_{\omega=0}^{N-1}z_\omega^{\rho_\omega}.
\end{equation}
Using eq.~\eqref{heat}, it is easy to prove that the $\mathbb{Q}$-function satisfies
\begin{equation}\label{Heat eq for Q}
0=\sum_{\omega}\nabla^{\mathfrak{q}}_\omega\log\mathbb{Q}-\frac{1}{2}\Delta_{\vec{z}}\log\mathbb{Q}+\frac{1}{2}\sum_\omega(\nabla^z_\omega\log\mathbb{Q})^2,
\end{equation}
with 
\begin{equation}
\sum_\omega\nabla_\omega^\mathfrak{q}=N\nabla^\mathfrak{q}+\vec{\rho}\cdot{\nabla}^{\vec{z}} .
\end{equation}
\bibliographystyle{utphys}
\bibliography{origami}

\end{document}